\documentclass[a4paper,fleqn]{cas-sc}

\usepackage[numbers]{natbib}
\usepackage{caption}
\usepackage{graphicx}
\usepackage{subfigure}
\usepackage{placeins}
\usepackage{multirow}
\usepackage{lineno}

\def\tsc#1{\csdef{#1}{\textsc{\lowercase{#1}}\xspace}}
\tsc{WGM}
\tsc{QE}
\tsc{EP}
\tsc{PMS}
\tsc{BEC}
\tsc{DE}

\begin{document}

\let\WriteBookmarks\relax
\def\floatpagepagefraction{1}
\def\textpagefraction{.001}

\shorttitle{Performance and pulse shape discrimination of glass scintillator SG101 for neutron detection}

\shortauthors{Yuhang Liu et~al.}

\title [mode = title]{Performance and pulse shape discrimination of glass scintillator SG101 for neutron detection}


%
\author[1]{Yuhang Liu}[type=editor,
                        auid=000,bioid=1,
                        prefix=,
                        role=,
                        orcid=]
\fnmark[1]


\credit{Writing– Original Draft, Investigation}

\address[1]{School of Physics, 
Sun Yat-sen University,No.135 West Xingang Road, 510275 Guangzhou , China
}
\author[1]{Fengpeng An}[type=editor,
                        auid=000,bioid=1,
                        prefix=,
                        role=,
                        orcid=]

\cormark[1]

\ead{anfp@mail.sysu.edu.cn}


\credit{Writing– Review, Conceptualization}
                        
\author[3]{Guang Luo}[%
   role=,
   suffix=,
   orcid=
   ]
\cormark[1]
\ead{luog7@mail2.sysu.edu.cn}

\credit{Writing– Review  , Visualization}

\address[2]{ Institut Franco-Chinois de l'Énergie Nucléaire , Sun Yat-sen University,
No. 2 Daxue Road,519082 Zhuhai, China
}

\author[1,2]{Wei Wang}
\ead{wangw223@mail.sysu.edu.cn}
\cormark[1]
\credit{Supervision}

\address[3]{School of Science, Sun Yat-sen University,
No. 66 Gongchang Road, 518107 Shenzhen, China}

\author[4]{Wei Wei}[%
   role=,
   suffix=,
   ]


\credit{Investigation}

\author[1]{Xuesong Zhang}[%
   role=,
   suffix=,
   ]
\credit{Discussion}
\author[1]{Dixiao Lu}[%
   role=,
   suffix=,
   ]


\credit{Discussion}
\author[1]{Xiaohao Yin}[%
   role=,
   suffix=,
   ]


\credit{Discussion}

\cortext[cor1]{Corresponding author}
\address[4]{Beijing Hoton Technology Co.Ltd.,
Hengye 7th Street, Tongzhou District,
100016 Beijing, China}



\begin{abstract}
We present a detailed characterization of the thermal neutron sensitive transparent glass scintillator SG101, benchmarked against the conventional LiF/ZnS:Ag-based scintillator EJ426. The detection efficiency,energy resolution, and pulse shape discrimination (PSD) performance of SG101 were evaluated under Am-Be neutron irradiation. When coupled with organic scintillators (EJ200 or EJ276),the SG101--EJ200 system achieves a figure-of-merit (FOM) of 3.81 for thermal neutron/gamma separation, while the SG101--EJ276 configuration resolves three distinct particle populations—gamma rays, fast neutrons, and thermal neutrons—with FOM values of 3.46 and 2.21, respectively.Correlation analysis reveals that the number of fast–thermal neutron coincidence events significantly exceeds the accidental background, and the count of $\gamma$–fast–thermal neutron triple-coincidence events is also far higher than the expected accidental rate, confirming significant physical correlations for both event types within a 100 \,\textmu s time window. These results demonstrate that SG101 is a promising candidate for applications requiring high-efficiency thermal neutron detection and precise event tagging coupling with a scintillator with PSD approach.
\end{abstract}

\begin{keywords}
Scintillators\sep Neutron detectors \sep Pulse shape discrimination \sep Coincidence \sep 
\end{keywords}


\begin{highlights}
\item 
\end{highlights}

\maketitle
\flushbottom

\section{Introduction}
\label{sec:intro}
Solid scintillators exhibit extensive application prospects in both high-energy physics and nuclear physics,including radiation monitoring, neutron scattering science, fusion research, nuclear security, and nuclear engineering.~\cite{Bernstein:2019hix,Brdar:2016swo,Bernstein:2001cz,Luo:2024xir}.
Given the widely use of \textsuperscript{6}Li-based materials in thermal neutron detection\cite{PROSPECT:2018hzo}, this study focuses on a novel \textsuperscript{6}Li-doped transparent glass scintillator (SG101)\cite{SG101} and comparatively evaluates its 
 performance against the established LiF/ZnS:Ag-based scintillator EJ426\cite{EljenEJ426,GAMAGE20141,Perrey:2021euc}. Under Am--Be neutron irradiation, the luminescent properties and detection efficiency of SG101 were systematically investigated. Furthermore, the particle discrimination capability and scintillation efficiency of hybrid detector configurations incorporating organic scintillators (EJ200 or EJ276)\cite{EljenEJ200,EljenEJ276} were characterized to comprehensively assess SG101’s potential as a high-performance thermal neutron-sensitive medium.

The remainder of this paper is organized as follows. Section~\ref{sec:pri} describes the experimental setup and measurement methodology. Section~\ref{sec:perform} presents a comparative study of the basic detection performance between SG101 and the conventional EJ426 scintillator. Section~\ref{sec:linear} demonstrates the energy response linearity of detector systems coupling SG101 with organic scintillators. Section~\ref{sec:PSD} details the pulse shape discrimination capabilities and the correlation analysis of fast-thermal neutron coincidence events. Finally, a summary of the key findings is provided in Section~\ref{sec:sum}.

\section{Experimental Setup and Methodology}
\label{sec:pri}

\begin{figure}[!htb]
\centering
\includegraphics[width=.9\textwidth]{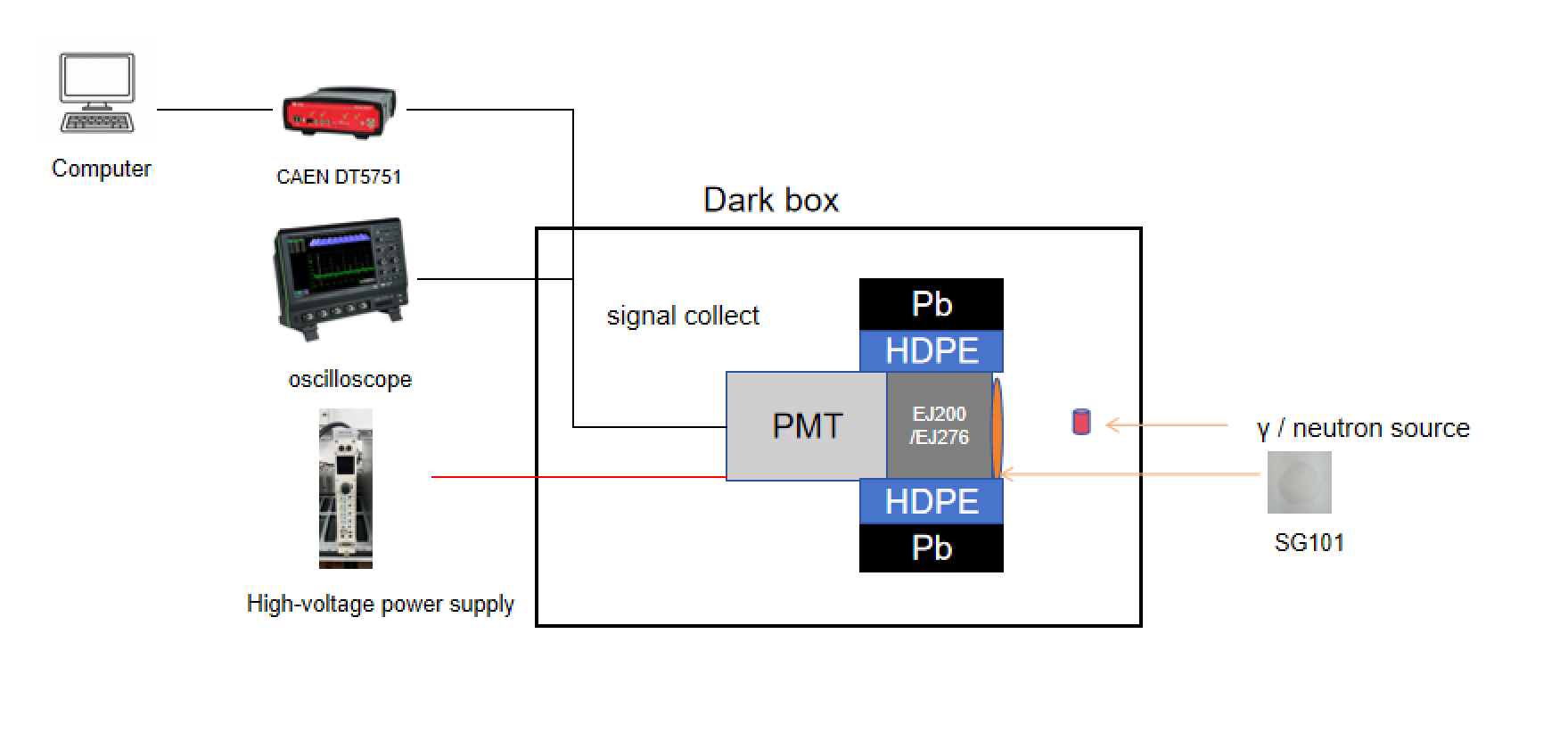}
\qquad
\includegraphics[width=.5\textwidth]{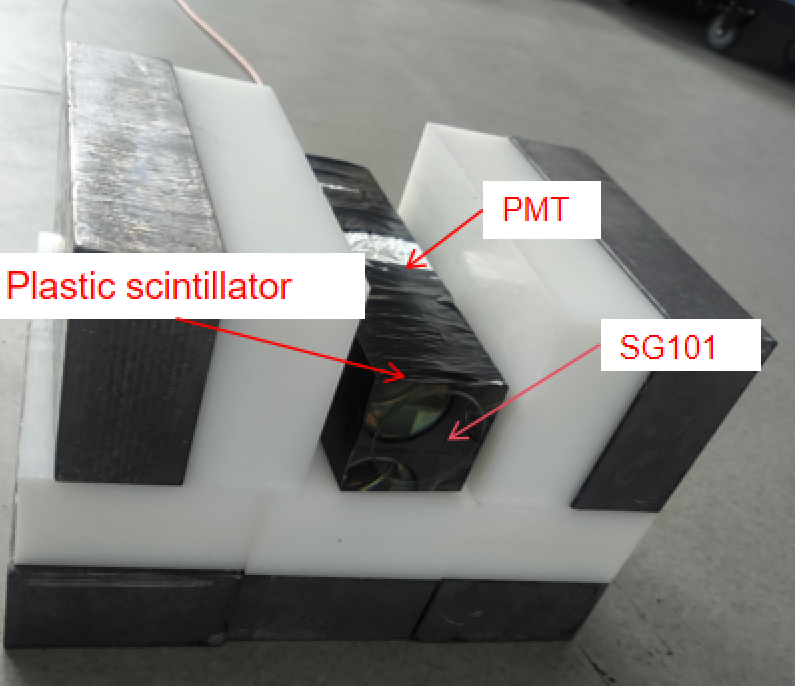}
\caption{Upper panel: Schematic of the experimental setup. The scintillator is optically coupled with silicone grease to a PMT on one end and to SG101 on the other end, all enclosed in a light-tightness box. The lead and HDPE bricks provide $\gamma$ and neutron shielding, respectively. Variable-thickness HDPE layers between SG101 and the neutron source enable neutron moderation. Data are acquired using either an oscilloscope or a CAEN DT5751 digitizer as needed. Lower panel: Photograph of the detector assembly.\label{fig:exp}}
\end{figure}

The experimental setup is shown in the upper panel of Figure~\ref{fig:exp},the lower panel is a photograph of the detector section. Signals were acquired either by a digital oscilloscope or by the CAEN DT5751 digitizer. The entire assembly was mounted on an external metal frame and enclosed in a light-tightness box  by wrapping it with thick black cloth
.

Optical coupling between the photomultiplier tube (PMT) and the scintillator cube was achieved using optical silicone grease. The detector was then wrapped with aluminum foil and secured with black tape, which not only improved light-tightness but also provided electromagnetic shielding.

The scintillator was fully surrounded by 
lead bricks to reduce the influence of environmental $\gamma$-ray background,and by high-density polyethylene(HDPE) bricks to shield against ambient neutrons.

\begin{figure}[!htb]
\centering
\includegraphics[width=0.29\textwidth]{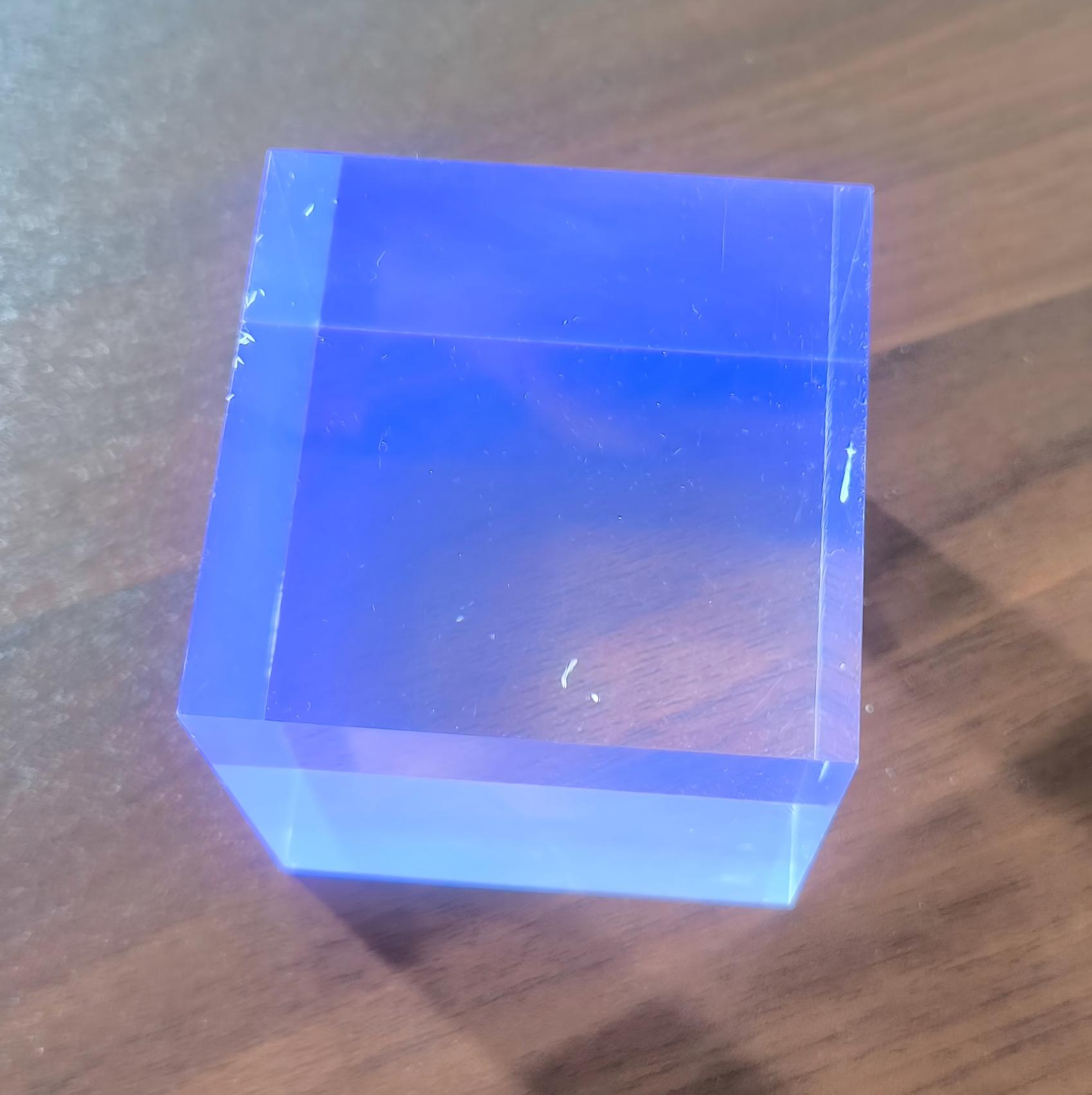}%
\hfill
\includegraphics[width=0.3\textwidth]{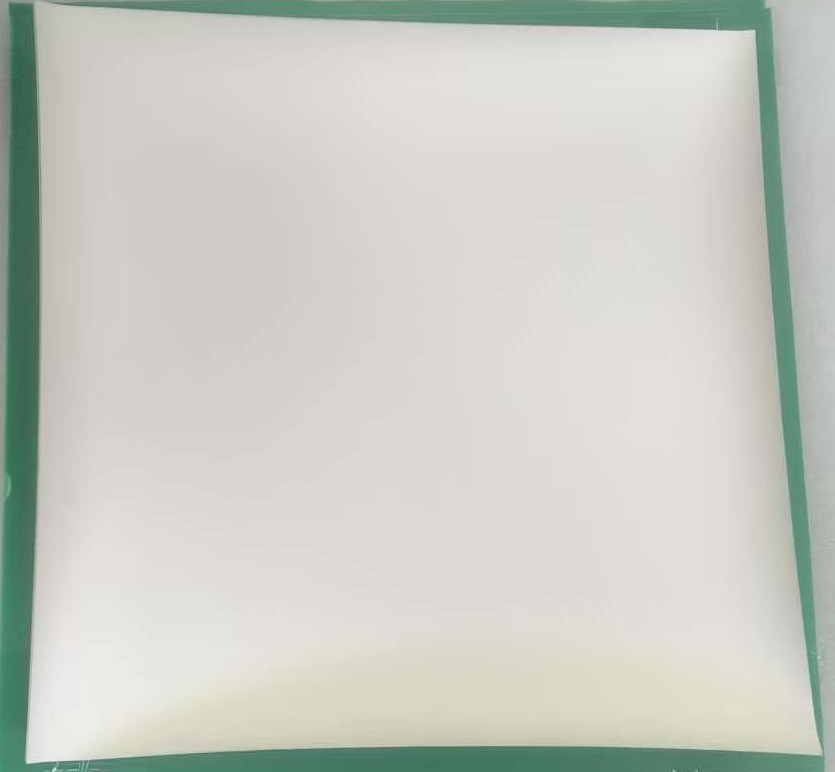}%
\hfill
\includegraphics[width=0.28\textwidth]{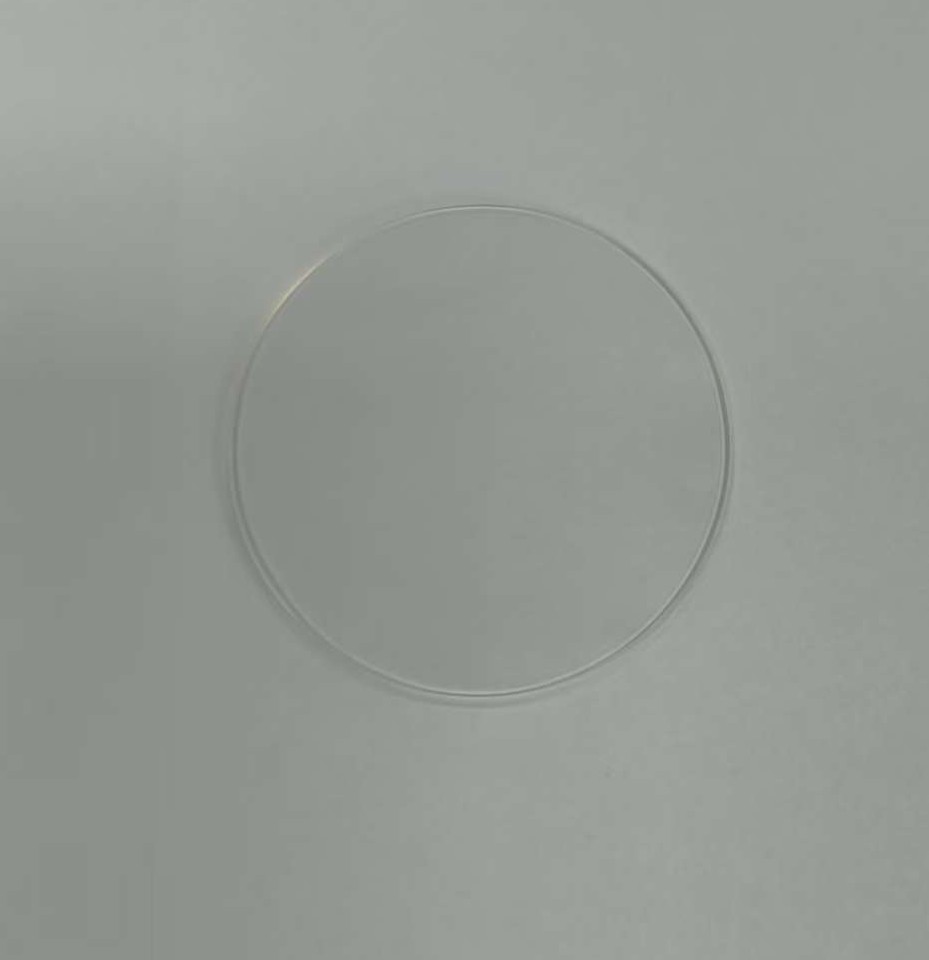}%
\caption{The left panel shows the EJ276 plastic scintillator,which has the same dimensions with EJ200.The central panel displays the EJ426 scintillator, while the right panel presents the SG101 scintillator.}
\label{fig:sc_swt}
\end{figure}

This study employed four scintillators: EJ200, EJ276, EJ426, and SG101\cite{EljenEJ200,EljenEJ276,EljenEJ426,SG101}.Figure~\ref{fig:sc_swt} shows photographs of various solid scintillators. EJ200 and EJ276 are cubic organic scintillators with an edge length of 6 cm, exhibiting excellent light yield. SG101 is a transparent circular glass disc with a diameter of 5 cm and a thickness of 1.07 mm. EJ426 is a 0.32 mm thick , 42 cm square non-transparent white sheet. Key parameters of these scintillators are summarized in Table~\ref{tab:1}.

Upon particle incidence, EJ276 exhibits markedly distinct waveform characters under $\gamma$-ray and neutron irradiation, enabling not only energy measurement of but also effective particle identification. In contrast, EJ200, despite having a comparable light yield, lacks the capability to discriminate between neutrons and $\gamma$ rays.

SG101 is a transparent glass scintillator fabricated by incorporating lithium oxide enriched in $^{6}$Li into a glass matrix (with a $^{6}$Li mass fraction of 7.5~wt\%), yielding a $^{6}$Li number density of approximately $7.4 \times 10^{21}~\mathrm{atoms/cm^{3}}$, and co-doped with a small amount of Ce$^{3+}$ ions~\cite{SG101}. This material primarily interacts with thermal neutrons via the \textsuperscript{6}Li$(n,\alpha)t$ nuclear reaction in Equation~\ref{eq:2.1}:
\begin{equation}
\label{eq:2.1}
    {}^{6}_{3}\text{Li} + \text{n} \rightarrow {}^{3}_{1}\text{H} + {}^{4}_{2}\text{He} + 4.78 \, \text{MeV}
\end{equation}
The reaction products, tritons (\textsuperscript{3}H) and alpha particles (\textsuperscript{4}He), deposit their energy in the glass, ionizing and exciting the Ce\textsuperscript{3+} luminescent centers~\cite{Jamieson:2017slg}. De-excitation of these centers yields visible light with an emission peak at 416 nm and a decay time of approximately 53ns~\cite{SG101}.Owing to its typically thin geometry, SG101 provides minimal stopping power for $\gamma$ rays. $\gamma$ rays deposit only a small fraction of their energy in the \textsuperscript{6}Li glass scintillator, producing fewer scintillation photons , this results in strong natural suppression of environmental $\gamma$-ray background\cite{Favalli:2025ngn}. These attributes make SG101 an excellent candidate material for thermal neutron detection.

EJ426 consists of a LiF/ZnS:Ag composite material and is inherently opaque; The $^{6}$Li concentration is $8.81 \times 10^{21}~\mathrm{atoms/cm^{3}}$.it likewise relies on the $^{6}$Li$(n,\alpha)t$ nuclear reaction in Equation~\ref{eq:2.1} to generate scintillation light.The ZnS:Ag composition in solid scintillator plays a role as a luminescent converter to enhance the overall light yield.\cite{Pino:2015dna}.

\begin{table}[!htb]
\centering
\caption{Basic parameters of the scintillator type.$^a$}
\label{tab:1}
\smallskip
\begin{tabular}{|l|lll|}
\hline
Scintillator type&Dimensions &Emission Peak (nm)&Decay time(ns)\\
\hline
EJ200 & Cube 6 cm $\times$ 6 cm $\times$ 6 cm & 425&2.1\\
EJ276 & Cube 6 cm $\times$ 6 cm $\times$ 6 cm & 450&$\gamma$:13;35;270 /n:13;59;460$^b$\\
EJ426 & Square Sheet 42 cm $\times$ 0.032 cm& 450& 200\\
SG101 & Circular sheet 5 cm $\times$ 0.107 cm & 416 & 53\\
\hline
\end{tabular}
\smallskip
\footnotesize
\\
$^a$Data from  technical data sheet ~\cite{EljenEJ200,EljenEJ276,EljenEJ426,SG101}.\\
$^b$Approximate mean decay times of the first three components (ns) for $\gamma$/neutron pulses.

\end{table}
The experiment employed an XP3232 PMT manufactured by Hainan Zhanchuang Photonics Technology Co., Ltd~\cite{Jiang:2022vnd}. The XP3232 is a cylindrical vacuum tube with a length of 11.2 cm and an outer diameter of approximately 5.1 cm. The PMT was directly coupled to the scintillator without an additional light guide. Optical silicone grease (Model SL612~\cite{SL612}, produced by Beijing Hoton Technology Co., Ltd.) was applied to fill the air gap between the scintillator and the PMT window. The PMT exhibits peak spectral sensitivity at 420 nm, which aligns well with the emission spectrum of the plastic scintillators used in this study.Through prior single-photoelectron gain calibration of this PMT, we determined that at a high voltage of 1200V, its gain is approximately $3 \times 10^{6}$, corresponding to a charge integral of the single-photoelectron signal of 24.15mV.ns~\cite{Liu2025}.

The oscilloscope used in this work is the HDO4104A high-resolution, high-precision instrument from Teledyne LeCroy~\cite{HDO4104A}, featuring a bandwidth of 1 GHz and a sampling rate of 10 GS/s. The signal acquisition system employs the DT5751 digitizer from CAEN, which offers a bandwidth of 500 MHz and a sampling rate of 1 GS/s. Compared to the oscilloscope, the DT5751 provides higher readout efficiency and built-in timestamp generation. The oscilloscope was utilized for precise waveform measurement and detailed waveform analysis of scintillator signals, while the DT5751 was employed to analyze particle discrimination capability and time-correlation characteristics of the scintillators.

Charged particles produce scintillation light with distinct decay times in different scintillators, resulting in significant differences in the temporal characteristics of their output waveforms. To exploit this property for particle identification, a short--long gate integration method was employed in this work\cite{Ranucci:1995un,ZAITSEVA2013747}. Specifically, the PSD value, denoted as Equation ~\ref{eq:3.1}:
\begin{equation}\label{eq:3.1}
\mathrm{PSD}_{\mathrm{value}} = 1 - \frac{Q_{\mathrm{short}}}{Q_{\mathrm{long}}}
\end{equation}

where \(Q_{\text{short}}\) and \(Q_{\text{long}}\) represent the integrated charge within a short time gate and the total integrated charge over the full pulse duration, respectively. In this experiment, the short integration window was set to 75\,ns and the long integration window to 800\,ns. For each recorded event, \(Q_{\text{short}}\) and \(Q_{\text{long}}\) were extracted from the digitized waveform, and the corresponding PSD value was computed.

\section{Comparative Performance of SG101 and EJ426}
\label{sec:perform}
\begin{figure}[!htb]
\centering
\includegraphics[width=.45\textwidth]{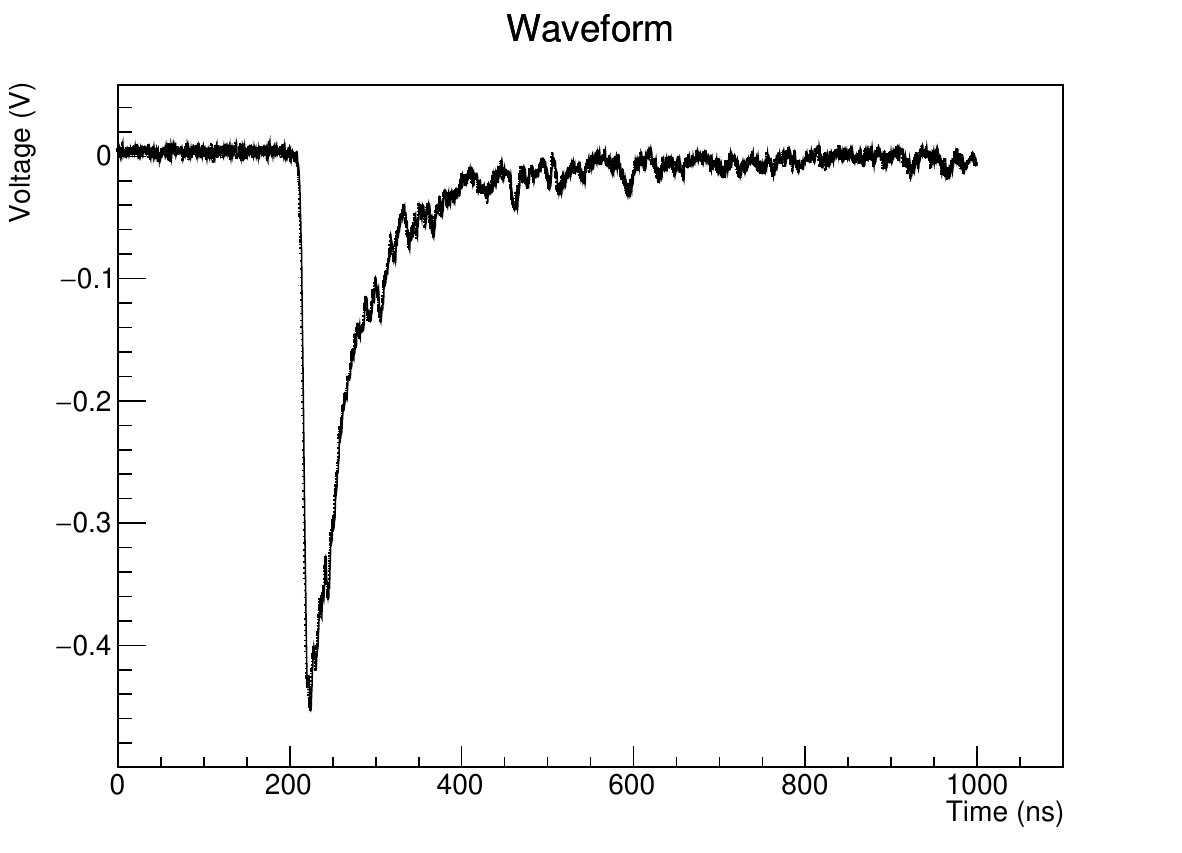}
\qquad
\includegraphics[width=.45\textwidth]{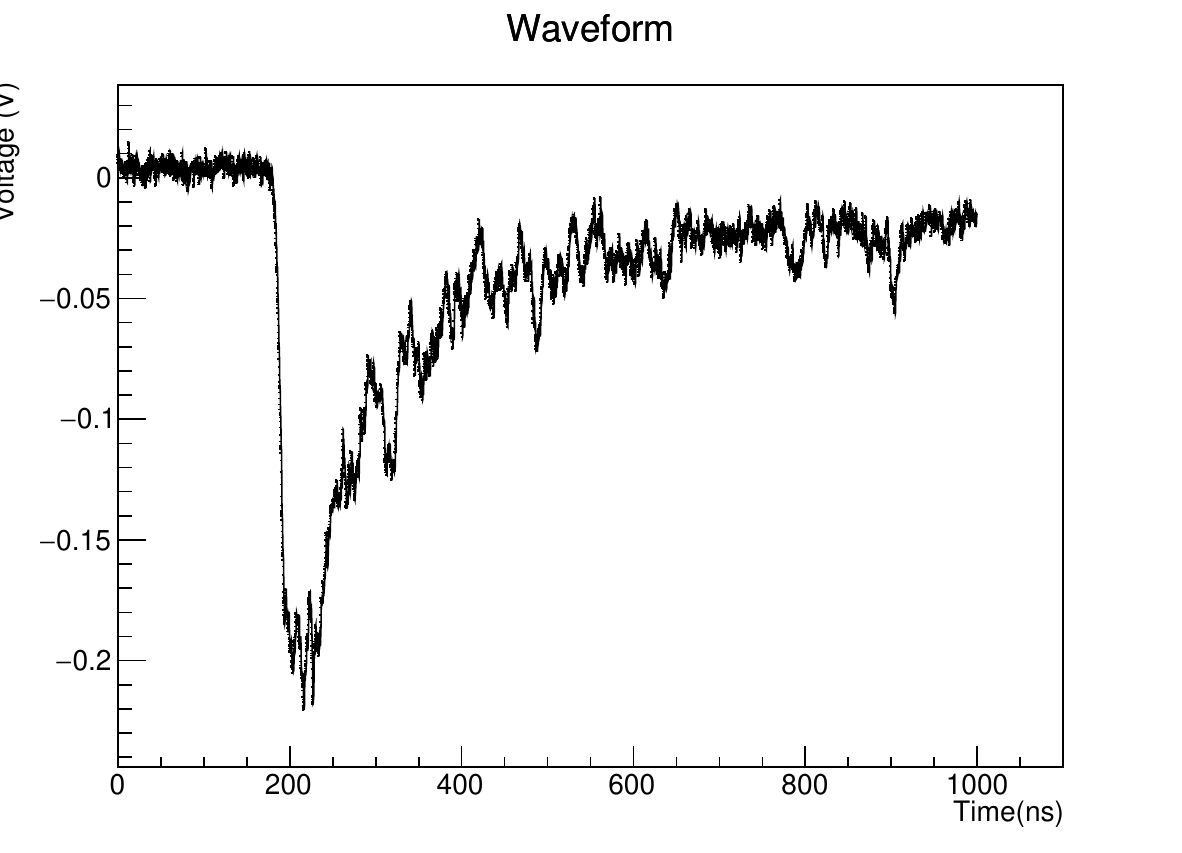}
\caption{Left panel:A single-pulse waveform sample from SG101 scintillator. Right panel:A single-pulse waveform sample from EJ426 scintillator which has many spikes in its falling edge compared with SG101 waveform.\label{fig:signal}}
\end{figure}

A stand alone test was done to compare the neutron detection performance of the two scintillators. The experimental setup is as follows: The EJ426 neutron screen and the SG101 scintillator were each directly coupled to a PMT, with an Am--Be neutron source positioned adjacent to the samples for irradiation. A typical single-event signal waveform is shown in Figure~\ref{fig:signal}.

EJ426 and SG101 are both \textsuperscript{6}Li-enriched thermal neutron-sensitive scintillators that utilize the \textsuperscript{6}Li($n$, $\alpha$)$t$ capture reaction to generate tritons (\textsuperscript{3}H) and alpha particles (\textsuperscript{4}He). 
 In EJ426, these charged particles deposit their energy in the ZnS:Ag phosphor, thereby inducing luminescence with a relatively broad scintillation pulse width of up to approximately 800\,ns. In contrast, SG101 produces light through the excitation of Ce\textsuperscript{3+} ions embedded in the glass matrix, resulting in a significantly narrower scintillation pulse.

\begin{figure}[!htb]
    \centering
\includegraphics[width=.45\textwidth]{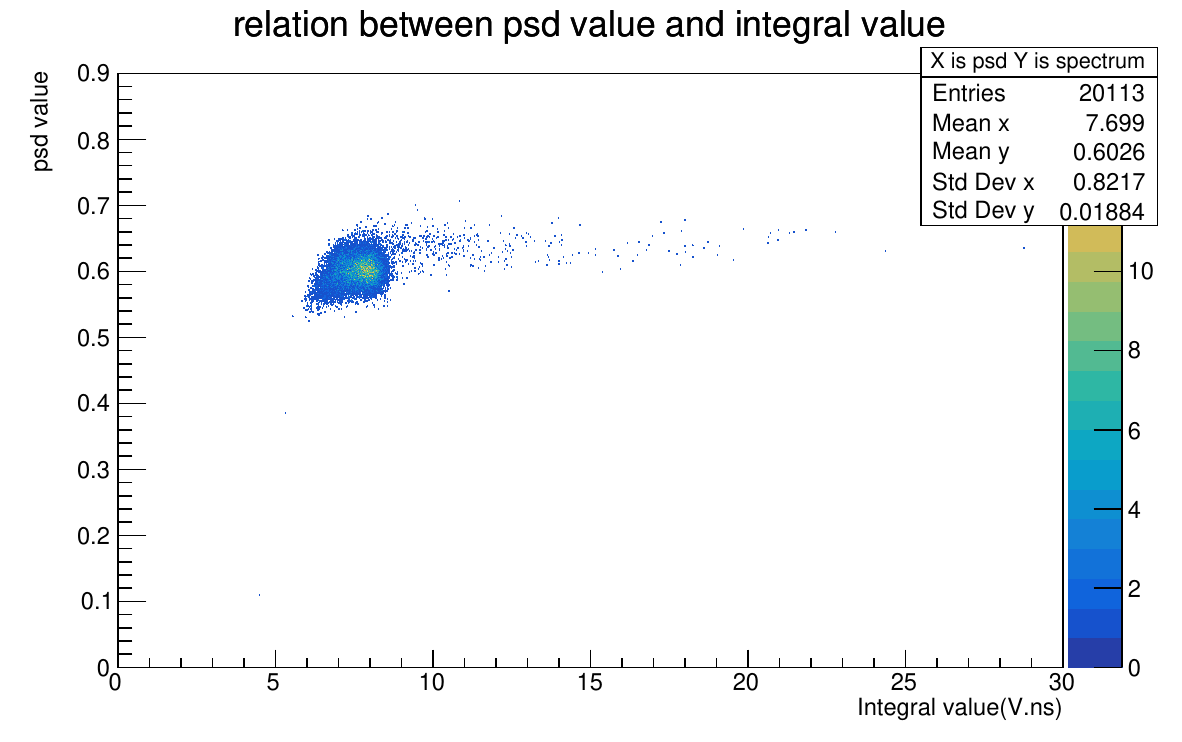}
\qquad
\includegraphics[width=.45\textwidth]{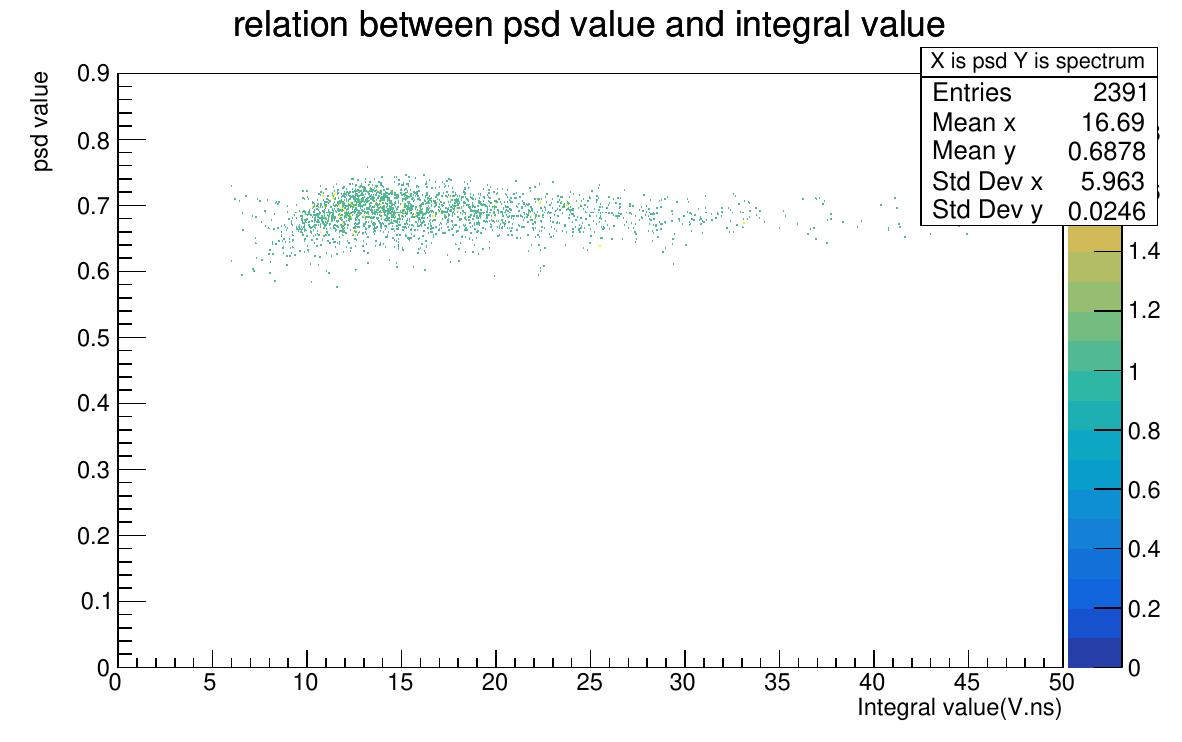}
    \caption{Left panel: 2D histogram of signal integral versus PSD values obtained using the SG101 detector; Right panel: corresponding distribution using the EJ426 detector. These figures are used to visually compare the differences in signal integral and PSD characteristics between the two different detector materials.}
    \label{fig:2dhist}
\end{figure}

In this setup, signals were acquired using oscilloscope. The data were analyzed by plotting a 2D histogram of PSD versus integrated signal amplitude, as shown in Figure~\ref{fig:2dhist}.EJ426 exhibits a broadened spectrum of integrated charge because the scintillation light yield from the reaction product \textsuperscript{4}He in ZnS:Ag has a broad statistical distribution~\cite{Pino:2015dna}.Resulting in a broad distribution of integrated signal amplitudes from 5 to 50 V·ns , with a mean of 16.69 V·ns and a standard deviation of 5.96 V·ns. In contrast, SG101 yields more stable waveforms and thus narrower distribution of the signal intergration, primarily within 5 to 25 V·ns, with a mean of 7.70 V·ns and a standard deviation of only 0.81 V·ns, indicating a more consistent and reproducible neutron response.

During the 10-minute signal acquisition period, HDPE layers of varying thicknesses were placed between the neutron source and the detector to thermalize the fast neutrons. Neutron signals detected by EJ426 and SG101 were statistically analyzed separately, with the resulting counts summarized in Table~\ref{tab:neutroncounts}.According to Section~\ref{sec:pri}, the $^{6}$Li number density in both SG101 and EJ426 is on the order of $10^{21}~\mathrm{atoms/cm^{3}}$.Although SG101 is thicker (0.107 cm) than EJ426 (0.032 cm), its higher optical transparency enables more efficient propagation of scintillation photons, resulting in significantly higher detection efficiency for SG101 over the same 10-minute period. 

\begin{table}[!htb]
    \caption{Neutron counts for EJ426 and SG101 after moderation by HDPE of different thicknesses.}
       \centering
    \begin{tabular}{|c|c|c|}
        \hline
        HDPE Thickness &Events counts (EJ426) & Events counts (SG101) \\
        \hline
        0 cm & 1057& 8275 \\
        1 cm & 1213& 8520 \\
        2 cm & 1815& 11620 \\
        3 cm & 2391& 20113 \\
        \hline
    \end{tabular}

    \label{tab:neutroncounts}
\end{table}
In summary, SG101 demonstrates a more concentrated amplitude distribution and greater signal stability, offering a better energy resolution in neutron detection compared to EJ426. Moreover, it achieves substantially higher neutron counts during the same acquisition period, underscoring its superior detection efficiency.

\section{Energy Linearity of Composite Detectors}
\label{sec:linear}
Plastic scintillators are mainly composed of low atomic number carbon and hydrogen atoms, which have a lower ability to block gamma rays. Therefore, energy calibration must be performed by reconstructing the Compton edge. This chapter aims to systematically verify the linear energy response of the detector:SG101 coupled with two plastic scintillators, EJ276 and EJ200, ensuring its accuracy in applications such as anti-neutrino detection. The verification is based on standard sources of $^{22}$Na, $^{137}$Cs, and $^{60}$Co, achieved by fitting the Compton edge and analyzing the linear relationship between energy and channel number.

The verification of energy response linearity follows the following procedure: First, the detector is irradiated with  gamma-ray sources---\(^{22}\)Na (511~keV and 1274.5~keV), \(^{137}\)Cs (661.7~keV), and \(^{60}\)Co (1173.2~keV and 1332.5~keV)---to establish the correspondence between incident gamma energy and detector response.

For each source, the acquired energy spectrum is analyzed by fitting a Gaussian function to the Compton edge region to determine its position. Using the previously calibrated single-photoelectron gain of the PMT---where one photoelectron corresponds to an integrated charge equivalent of 24.15~mV$\cdot$ns\cite{Liu2025}---the Compton edge positions are converted into the corresponding number of photoelectron (p.e.). This yields a set of p.e. values as a function of gamma-ray energy, enabling quantitative assessment of the detector's energy linearity.

\begin{figure}[ht]
    \centering
    \begin{minipage}[t]{0.68\textwidth}
        \vspace{0pt}  
        \includegraphics[width=\linewidth]{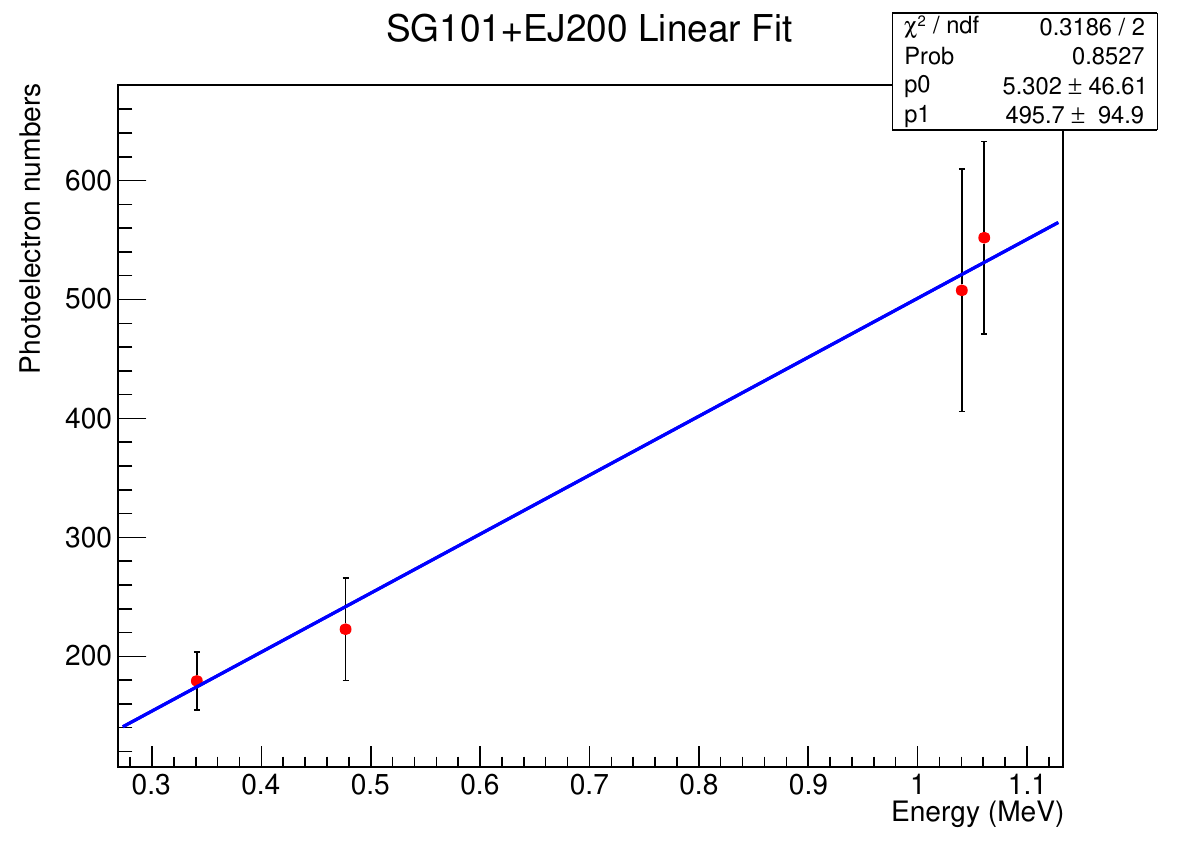}
    \end{minipage}
    \hfill
    \begin{minipage}[t]{0.3\textwidth}
        \vspace{0pt}  
        \centering
        \includegraphics[width=\linewidth]{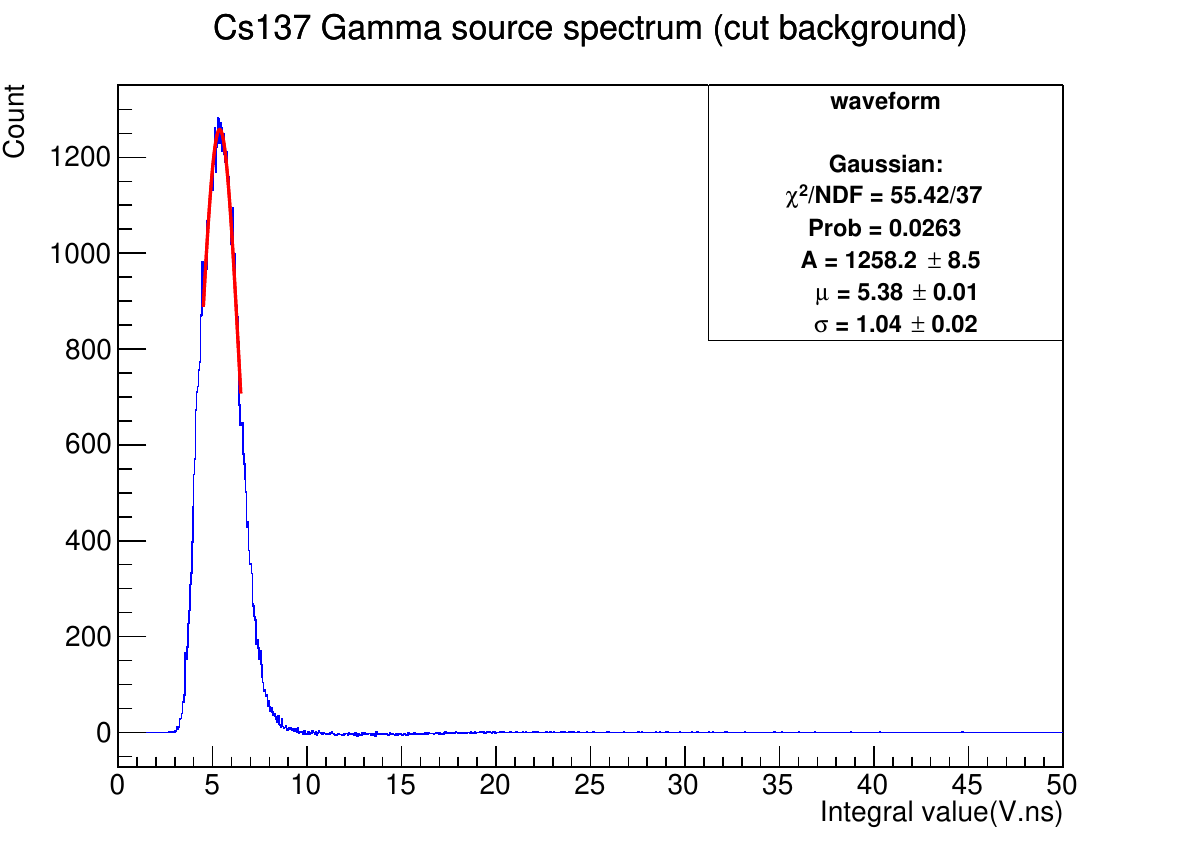} \\[1em]
        \includegraphics[width=\linewidth]{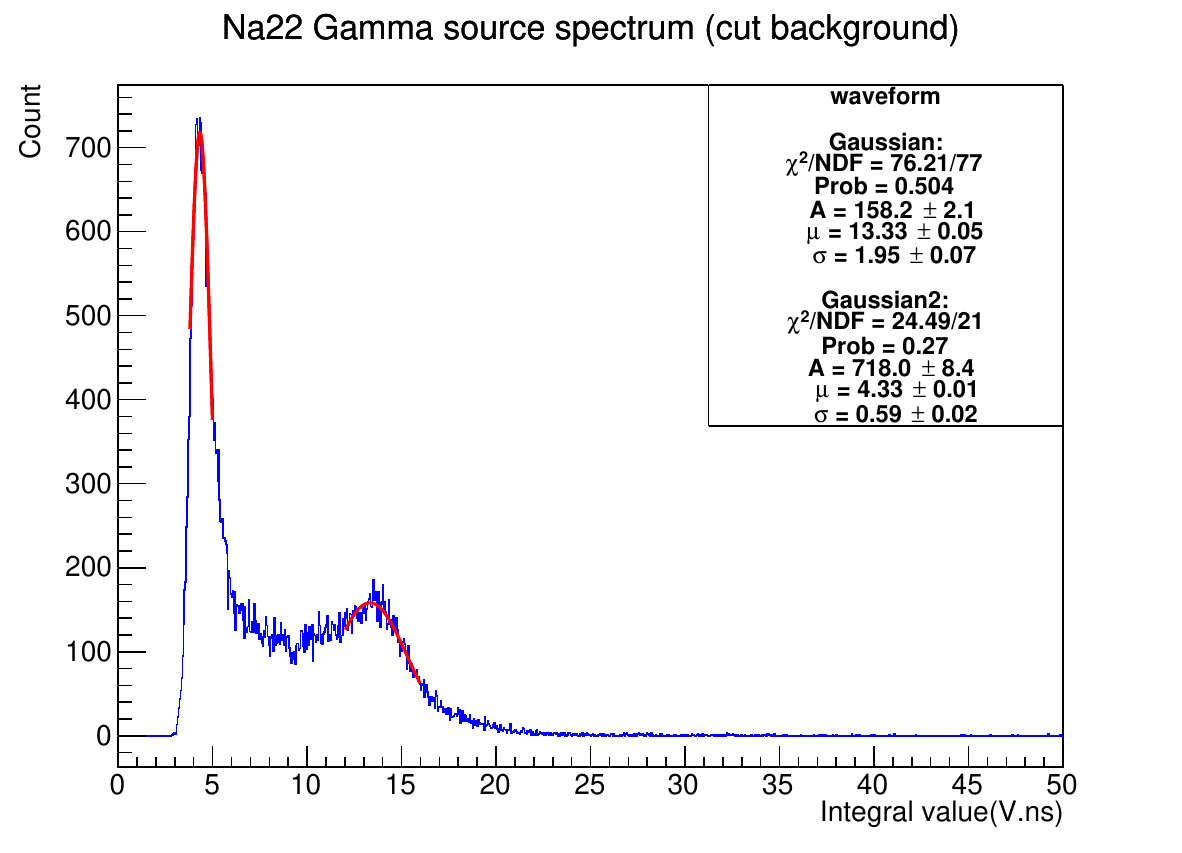} \\[1em]
        \includegraphics[width=\linewidth]{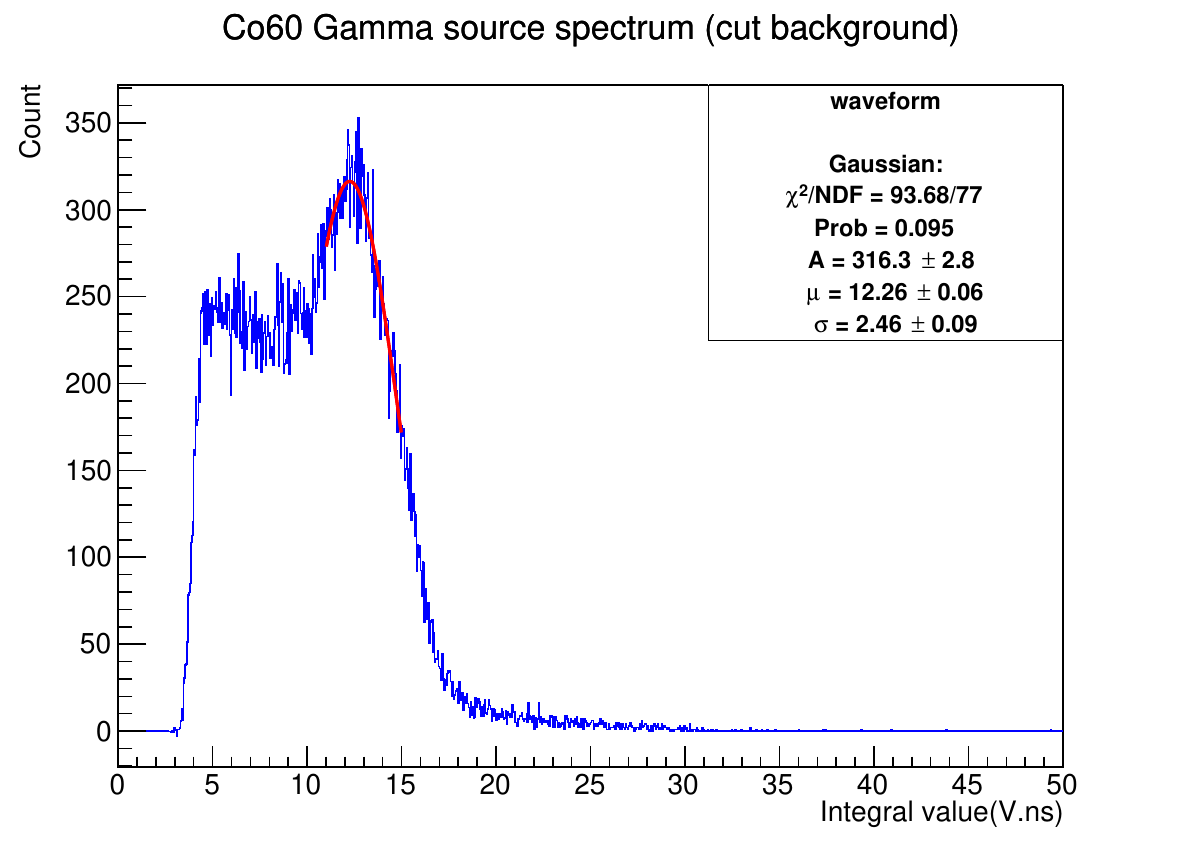}
    \end{minipage}
    \caption{Left: Linear fit results using data from the EJ200 + SG101 detector combination. Right: Gaussian fits to the Compton edge energies of \textsuperscript{137}Cs (0.477MeV), \textsuperscript{22}Na (0.341MeV and 1.061MeV), and \textsuperscript{60}Co (0.9632MeV and 1.1182MeV). Given that the two \textsuperscript{60}Co peaks are too close for accurate double-Gaussian fitting; therefore, their average value was used. The left figure shows a linear fit based on the mean and standard deviation values from the Gaussian fits, converted into photoelectron numbers.}
    \label{fig:200lbl_linear}
\end{figure}

Figure~\ref{fig:200lbl_linear} presents the energy linearity verification for the EJ200+SG101 detector. In the right panel, Compton edge positions corresponding to \(^{22}\text{Na}\), \(^{137}\text{Cs}\), and \(^{60}\text{Co}\) gamma-ray sources are extracted via Gaussian fits, with uncertainties dominated by statistical fluctuations. 

The left panel displays the resulting linear relationship between gamma-ray energy and photoelectron number, where data points correspond to the fitted Gaussian means (\(\mu\)) and error bars represent the associated standard deviations (\(\sigma\)).
The linear fit is expressed as
\[
\text{p.e.} = p1\times E + p0,
\]
where \(E\) denotes the incident gamma-ray energy (in keV or MeV), and \(\text{p.e.}\) is the corresponding number of photoelectrons. The offset parameter \(p0\) primarily accounts for systematic effects such as trigger threshold settings and baseline drift.The slope \(p1\) represents the detector’s photoelectron yield per unit energy (i.e., number of photoelectrons per MeV).The energy scale of the system is $495.7\,\mathrm{p.e./MeV}$.

\begin{figure}[ht]
    \centering
    \begin{minipage}[t]{0.66\textwidth}
        \vspace{0pt}  
        \includegraphics[width=\linewidth]{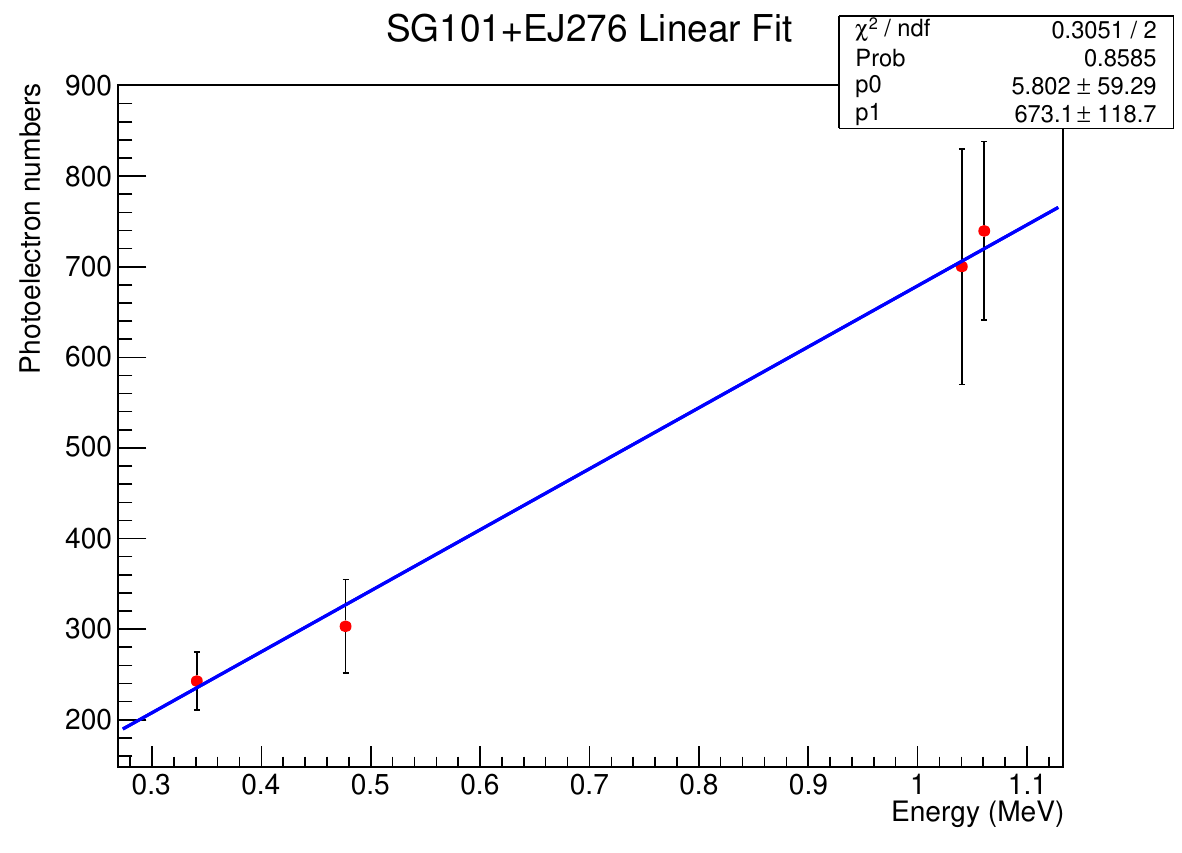}
    \end{minipage}
    \hfill
    \begin{minipage}[t]{0.3\textwidth}
        \vspace{0pt}  
        \centering
        \includegraphics[width=\linewidth]{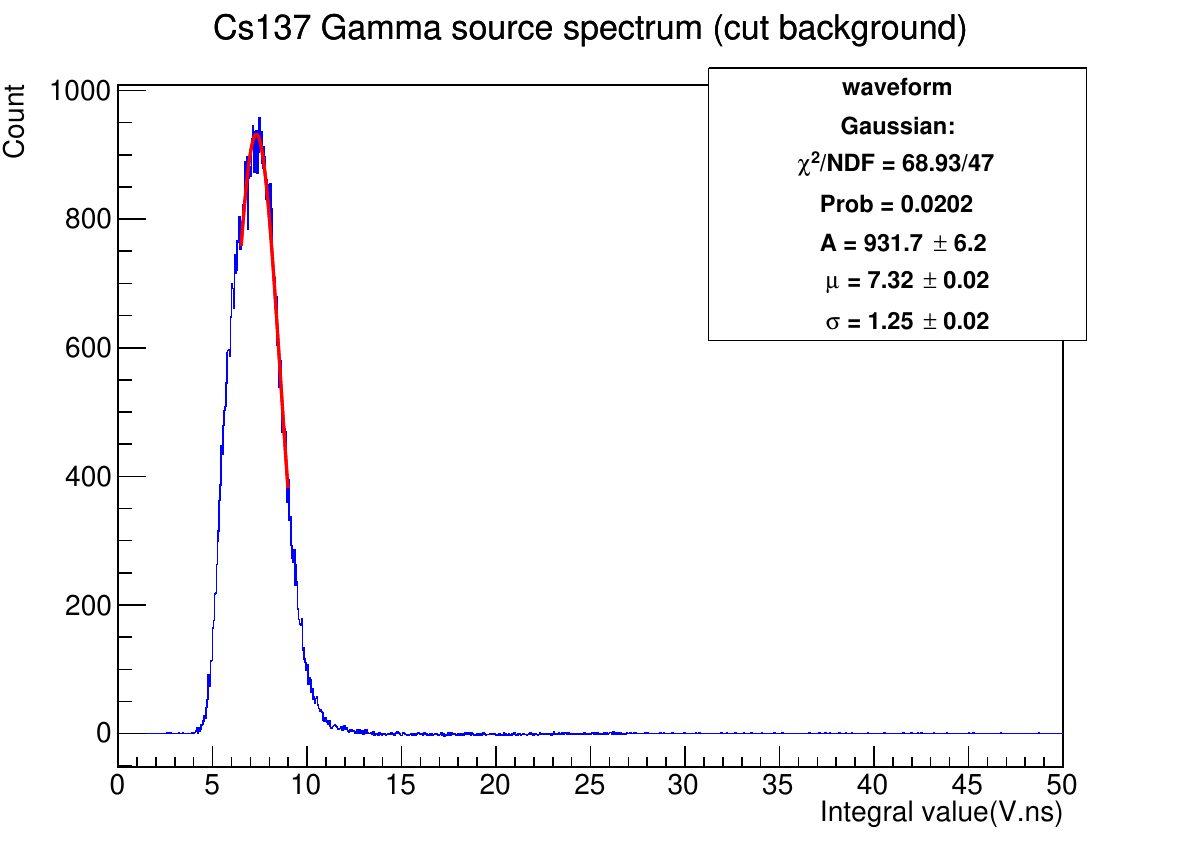} \\[1em]
        \includegraphics[width=\linewidth]{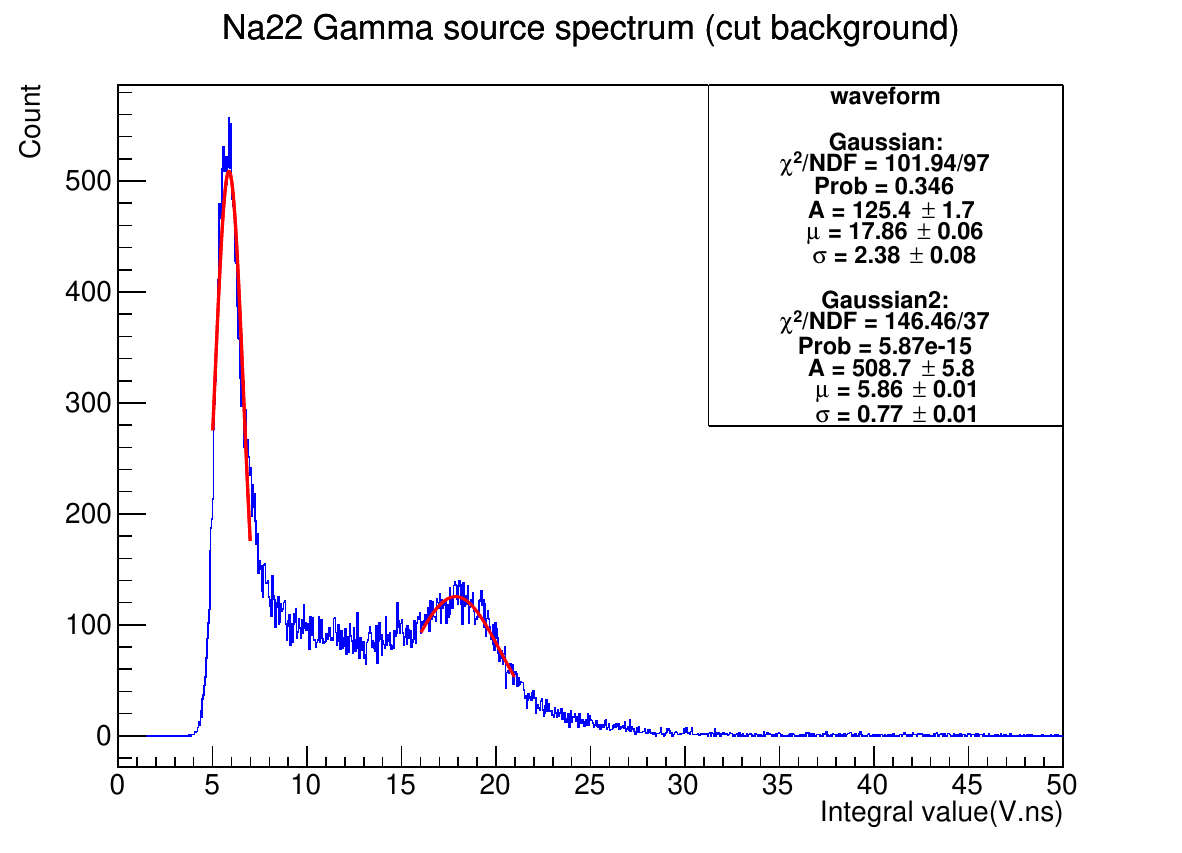} \\[1em]
        \includegraphics[width=\linewidth]{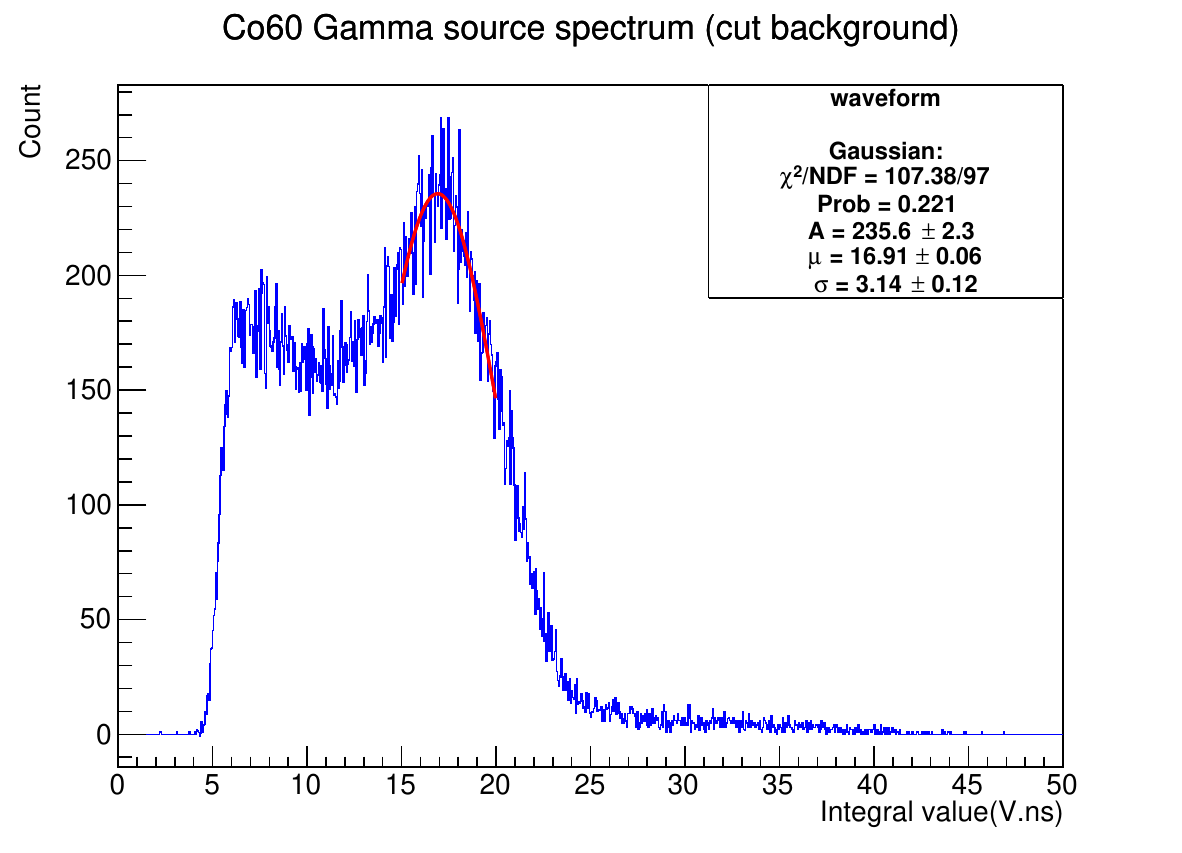}
    \end{minipage}
    \caption{Left: Linear fit results using the EJ276 + SG101 detector setup. Right: Gaussian fits to the Compton edge energies of \textsuperscript{137}Cs (0.477MeV), \textsuperscript{22}Na (0.341MeV and 1.061MeV), and \textsuperscript{60}Co (average of 0.9632MeV and 1.1182MeV due to close proximity). The left panel shows a linear fit based on these Gaussian parameters, converted into photoelectron numbers.}
    \label{fig:276lbl_linear}
\end{figure}

A similar analysis is performed for the EJ276+SG101 detector, as shown in Figure~\ref{fig:276lbl_linear}. The right panel illustrates the identified Compton edges from the same set of radioactive sources, again limited by counting statistics. After converting these edge positions into equivalent photoelectrons yields, a linear fit is carried out (left panel), using \(\mu\) as the central value and \(\sigma\) as the uncertainty for each point..The energy scale of the system is $673.1\,\mathrm{p.e./MeV}$.

The linearity analysis demonstrates that both the SG101+EJ276 and SG101+EJ200 detectors exhibit good energy linearity over the tested energy range (approximately 0.3MeV to 1.1MeV). Among them, the EJ276 detector coupled with optical silicone grease shows a higher photoelectron yield, consistent with previous experimental results\cite{Liu2025}. This energy response verification experiment conclusively confirms the linearity of the organic scintillators used, providing a reliable foundation for their application in subsequent energy reconstruction algorithms.

\section{Pulse Shape Discrimination and Coincidence Analysis}
\label{sec:PSD}

\begin{figure}[!htb]
    \centering
\includegraphics[width=.45\textwidth]{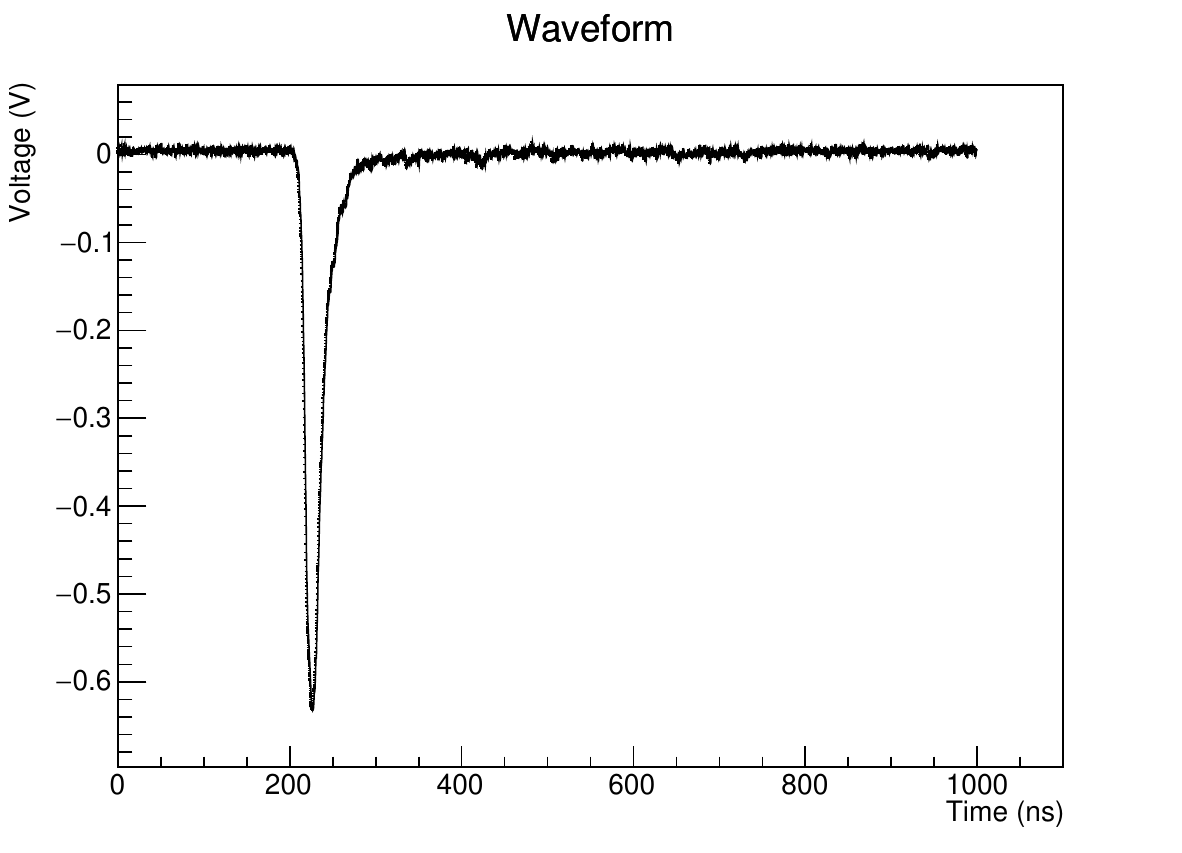}
\qquad
\includegraphics[width=.45\textwidth]{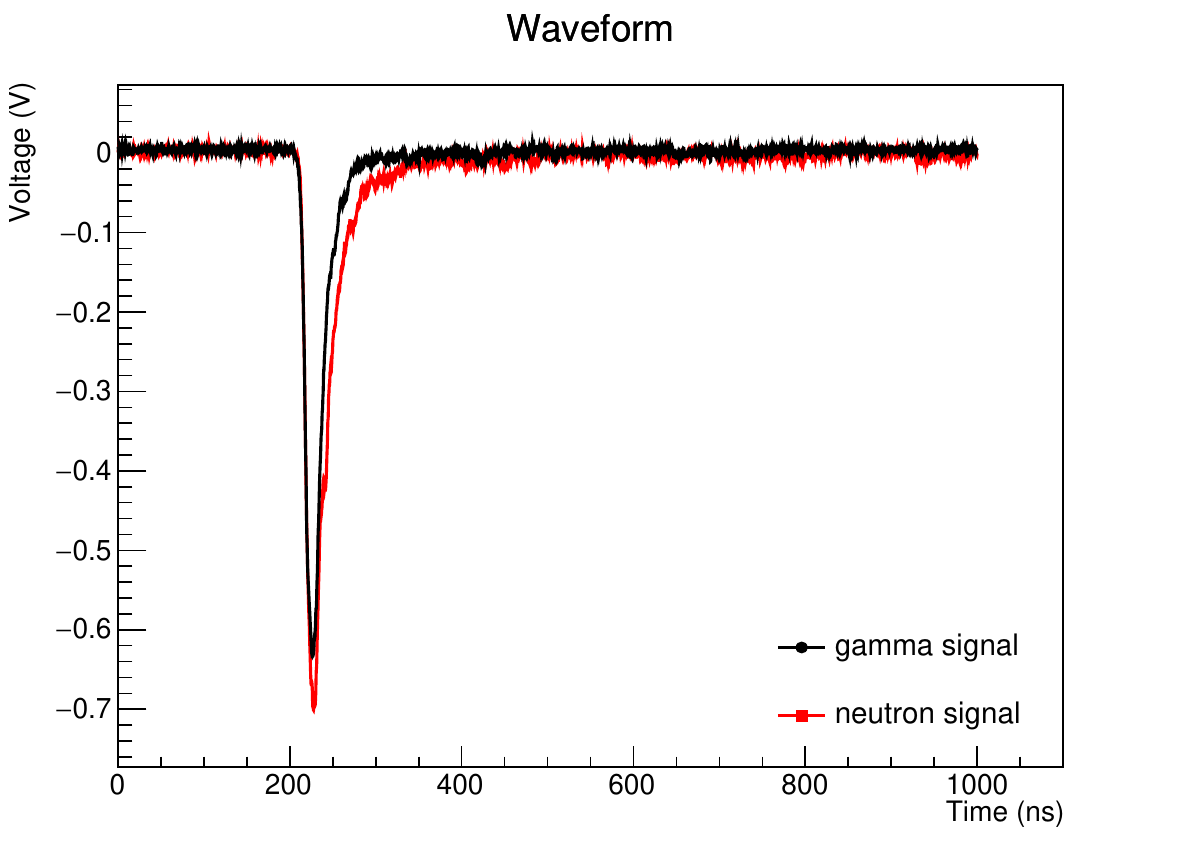}
    \caption{Left panel: single gamma pulse in EJ200. Right panel: single neutron pulse and signal gamma in EJ276.}
    \label{fig:276200Waveform}
\end{figure}

Figure~\ref{fig:276200Waveform} shows the typical single-event pulse waveforms from the two plastic scintillators. The EJ200 signal cannot distinguish between fast neutrons and $\gamma$ rays, whereas EJ276 effectively discriminates fast neutron events from $\gamma$-ray events. Notably, the pulse shapes of both plastic scintillators are markedly distinct from the neutron capture signals produced by the two thermal neutron-sensitive solid scintillators.

To evaluate particle discrimination capability, the SG101 scintillator was optically coupled with two types of plastic scintillators (EJ200 and EJ276), and PSD analysis was performed based on their output signals.The experimental setup is shown in Figure~\ref{fig:exp}. As shown in Figure~\ref{fig:276200Waveform}, the plastic scintillators exhibit pulse shapes that are markedly different from that of SG101(Figure~\ref{fig:signal}). This distinct temporal characteristic enables effective separation of neutron and gamma-ray events in the coupled system, yielding a well-resolved PSD distribution.

To further quantify the discrimination performance, the PSD spectrum was plotted, fitting with double-Gaussian functions for the neutron and gamma peaks,and the FOM was calculated accordingly to objectively characterize the $n/\gamma$ discrimination capability of this scintillator combination.The FOM, defined in Equation~\ref{eq:fom}, quantifies this separation:

\begin{equation}
\label{eq:fom}
\mathrm{FOM} = \frac{|\mathrm{S}_{n} - \mathrm{S}_{\gamma}|}{\mathrm{FWHM}_{n} + \mathrm{FWHM}_{\gamma}}
\end{equation}
Here, \( S_n \) and \( S_\gamma \) are the medians of the two Gaussian distributions, and FWHM denotes the full width at half maximum of the Gaussian peaks. The narrower the peaks and the larger the separation between their medians, the stronger the discrimination capability is. When the FOM value exceeds 1.27, the peak separation is greater than \( 3\sigma \), indicating that PSD can very effectively distinguish waveform signals from different particles~\cite{Winyard:1971}.

\begin{figure}[!htb]
  \centering
\includegraphics[width=.45\textwidth]{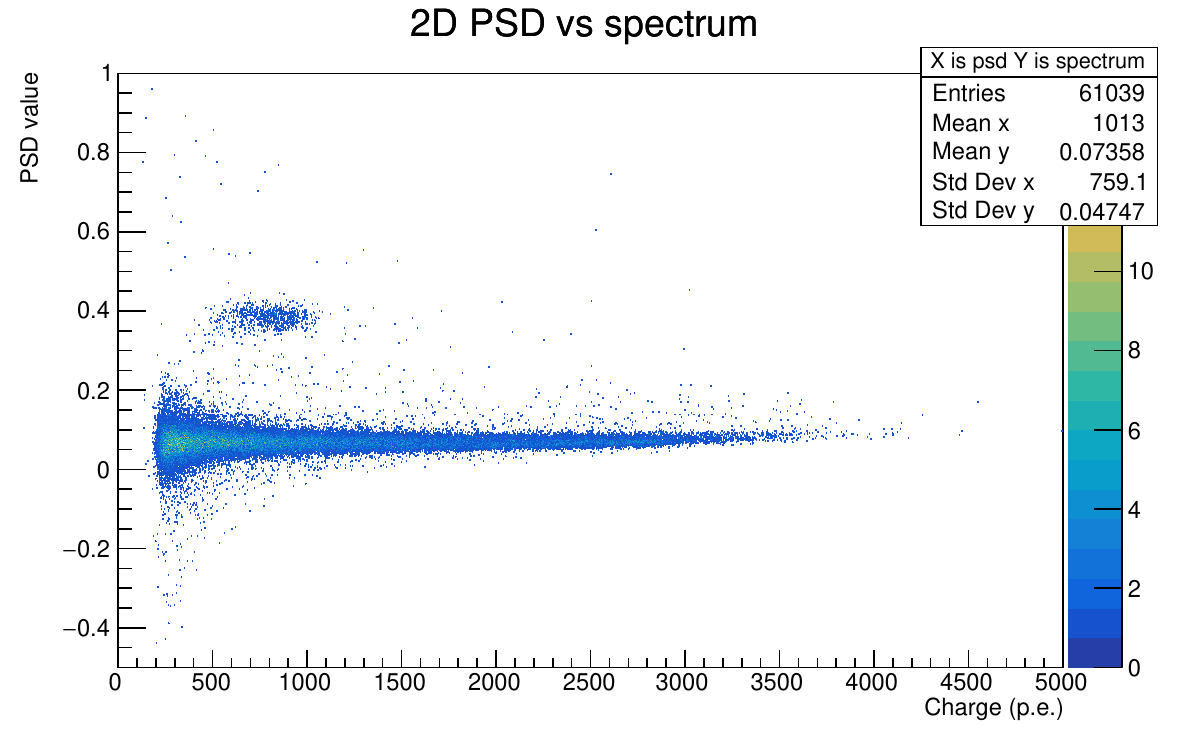}
\qquad
\includegraphics[width=.40\textwidth]{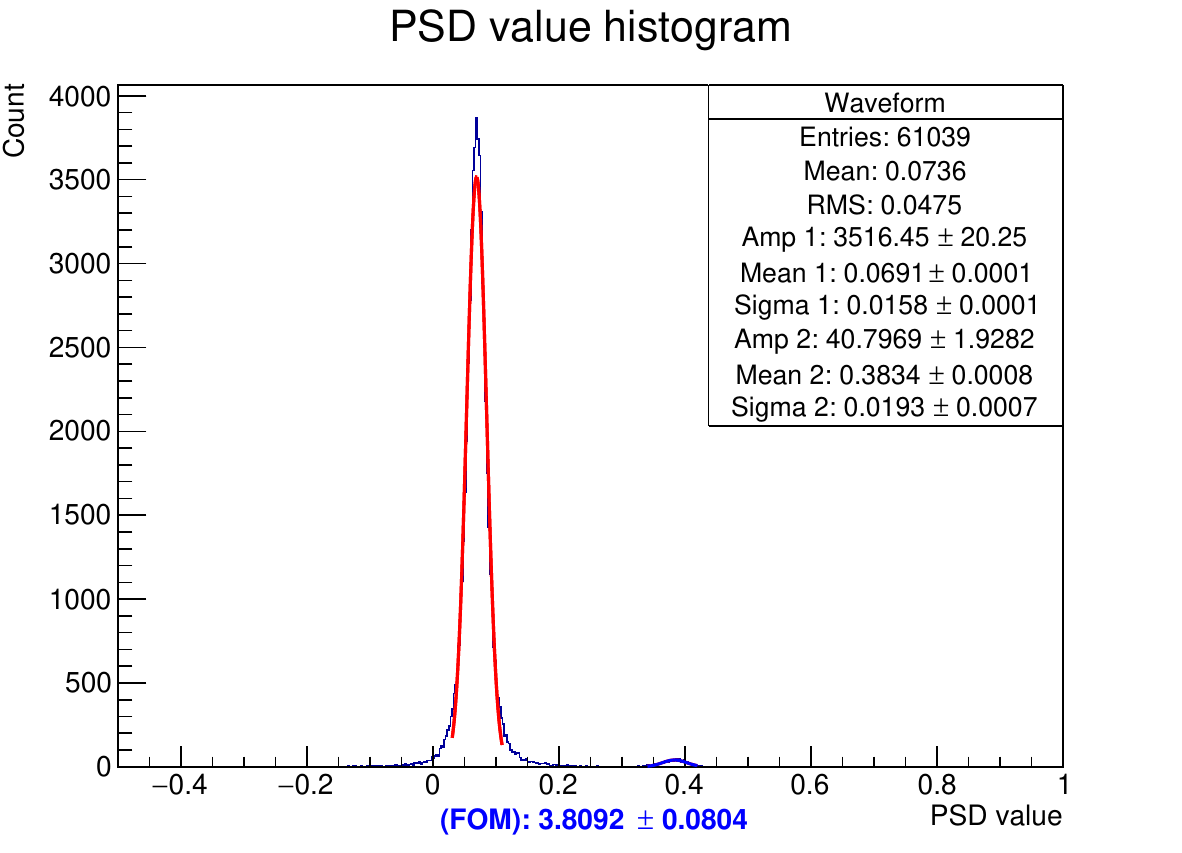}
    \caption{Left panel: 2D histogram of signal integral versus PSD values obtained using the SG101+EJ200 detector;Right panel: One-dimensional histogram of PSD values, fitted with a double-Gaussian function; the corresponding FOM value is listed below the histogram.}
    \label{fig:lbl200psd}
\end{figure}

Figure ~\ref{fig:lbl200psd} left panel shows the 2D PSD distribution of the coupled SG101–EJ200 detector under irradiation from an Am–Be neutron source. The horizontal axis represents energy, calibrated in units of photoelectron number, while the vertical axis shows the PSD value. Signals with a PSD value around 0.4 correspond to the slow scintillation component resulting from thermal neutron capture in SG101, whereas signals in the PSD range of 0–0.2 primarily originate from fast scintillation components induced by particle interactions in EJ200 (e.g., $\gamma$ rays or fast neutrons). Figure \ref{fig:lbl200psd} right panel displays the corresponding one-dimensional PSD spectrum; Gaussian fitting yields an FOM of 3.81. The separation between the thermal neutron and gamma-ray peaks exceeds five standard deviations (5$\sigma$), demonstrating that the SG101–EJ200 composite scintillator system exhibits excellent discrimination capability between thermal neutrons and $\gamma$ rays.

\begin{figure}[!htb]
    \centering
\includegraphics[width=.45\textwidth]{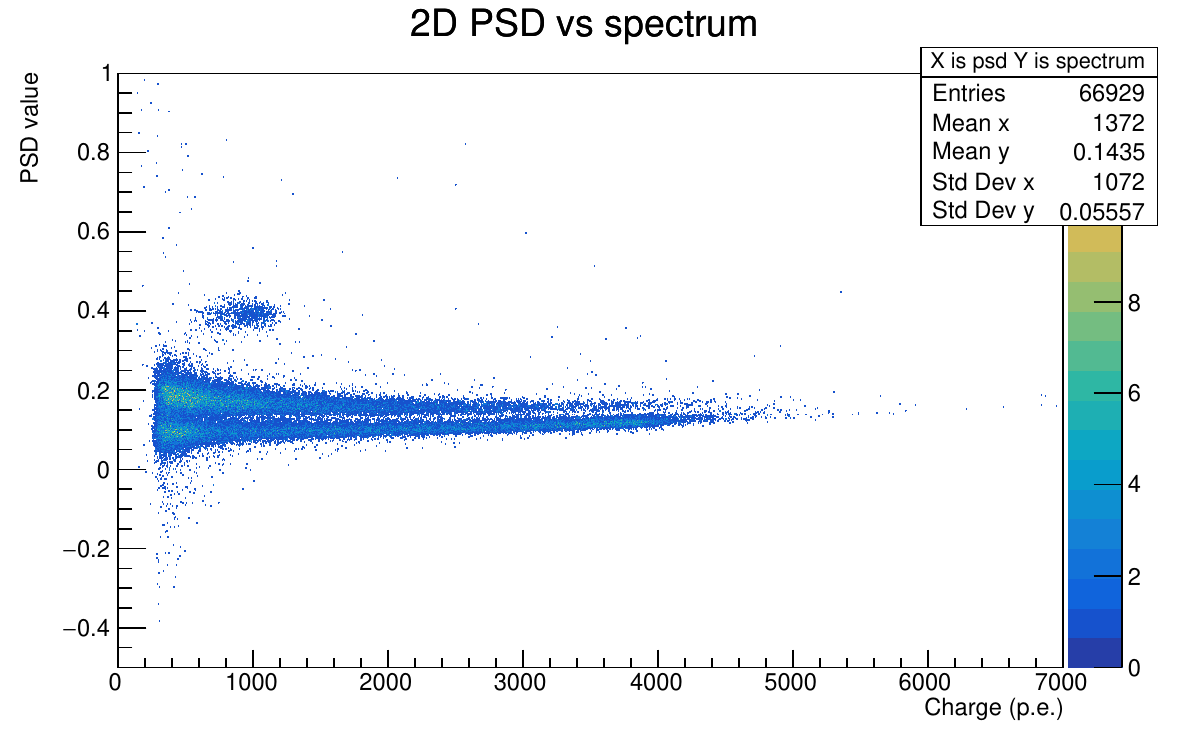}
\qquad
\includegraphics[width=.40\textwidth]{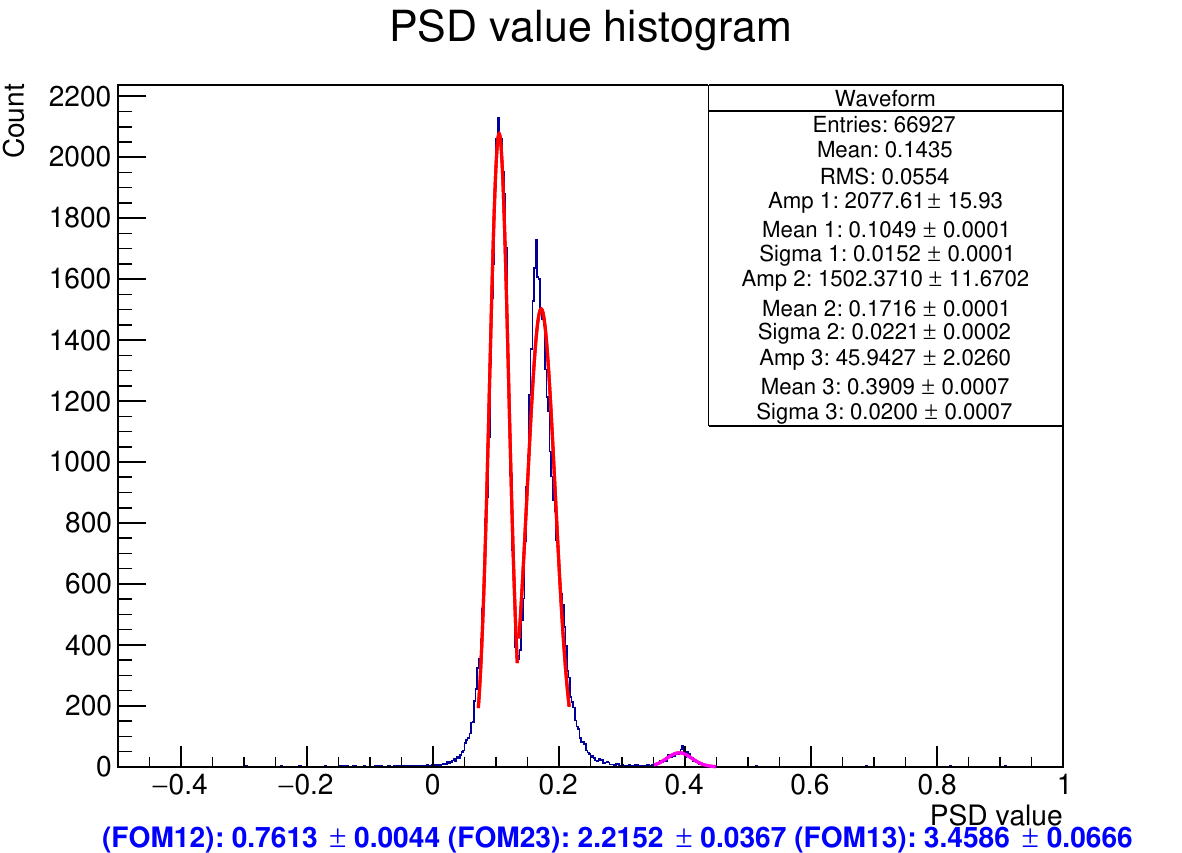}
    \caption{Left panel: 2D histogram of signal integral versus PSD values obtained using the SG101+EJ276 detector;Right panel: One-dimensional PSD histogram fitted with three Gaussian peaks; the figure lists FOM12, FOM13, and FOM23 below, representing the FOM values for peak pairs 1–2, 1–3, and 2–3, respectively.}
    \label{fig:lbl276psd}
\end{figure}

Figure ~\ref{fig:276200Waveform} right panel shows the pulse waveform generated by the EJ276 scintillator in response to a single neutron. Compared to gamma-ray signals, neutron-induced pulses exhibit slightly longer decay times and broader pulse widths. Consequently, when coupled with the SG101 scintillator, fast neutron signals can be clearly distinguished from gamma-ray and thermal neutron signals. The system's PSD spectrum reveals three distinct peaks in Figure ~\ref{fig:lbl276psd}.The PSD spectrum was fitted with three Gaussian peaks, yielding the mean and standard deviation ($\sigma$) for each peak. The FOM was calculated for each pair of peaks based on their separation. The FOM values between the thermal neutron peak and the $\gamma$ peak, and between the fast neutron peak and the thermal neutron peak, are 3.46(8.1$\sigma$)and 2.21(5.2$\sigma$), respectively. The thermal neutrons can be clearly separated by SG101 with fast neutrons or $\gamma$ signals at $5\sigma$ confidence level.The coupling of SG101 with EJ276 also preserves EJ276's excellent discrimination capability between fast neutrons and gamma rays. Previous experiments have shown that, for gamma-equivalent energies above 1\,MeV, the FOM of EJ276 remains consistently above 1.27 across all energy ranges.This indicates that fast neutrons and gamma-ray signals can be effectively discriminated at the $3\sigma$ confidence level\cite{Liu2025}.

\begin{figure}[!htb]
    \centering
\includegraphics[width=.43\textwidth]{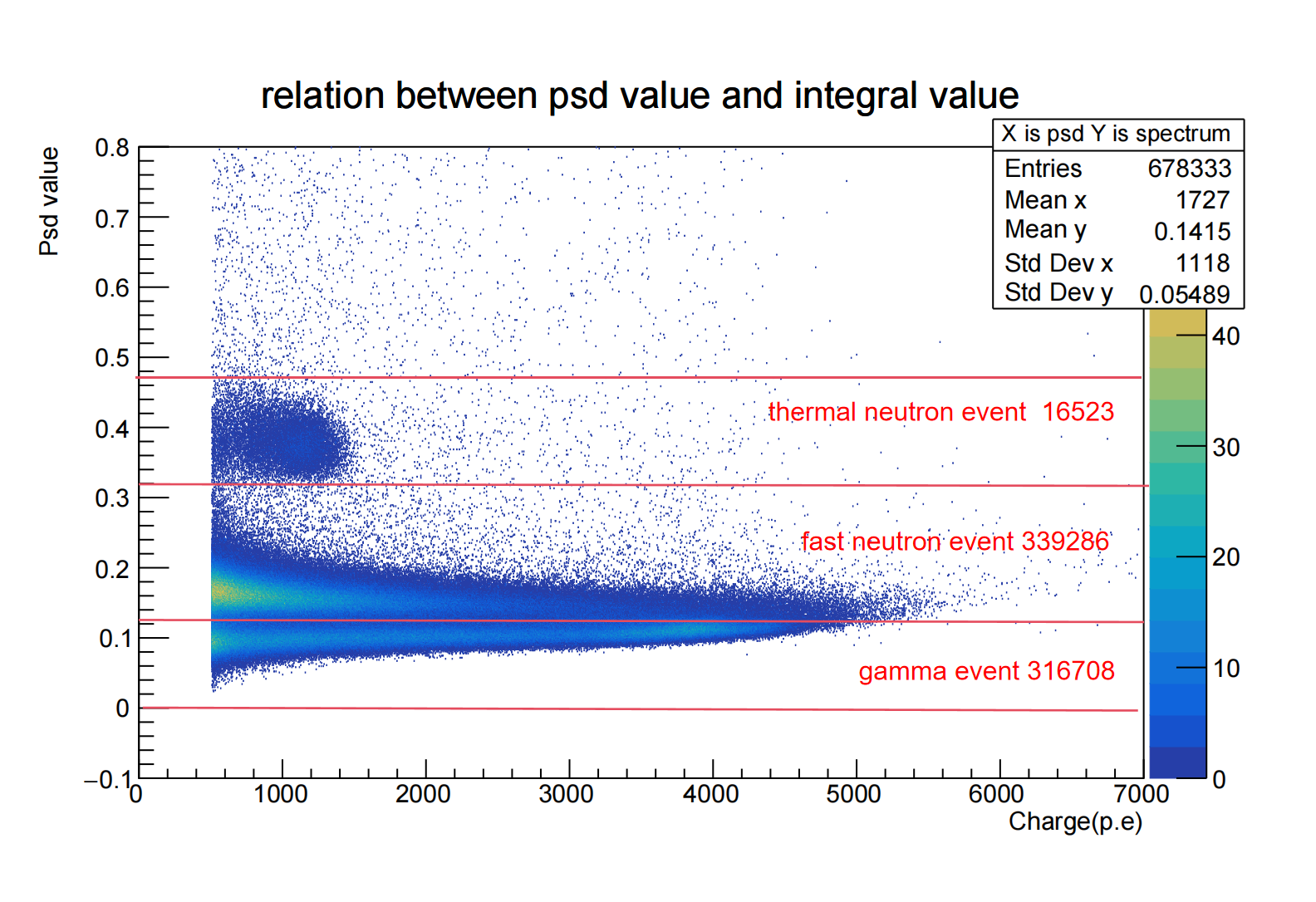}
\qquad
\includegraphics[width=.5\textwidth]{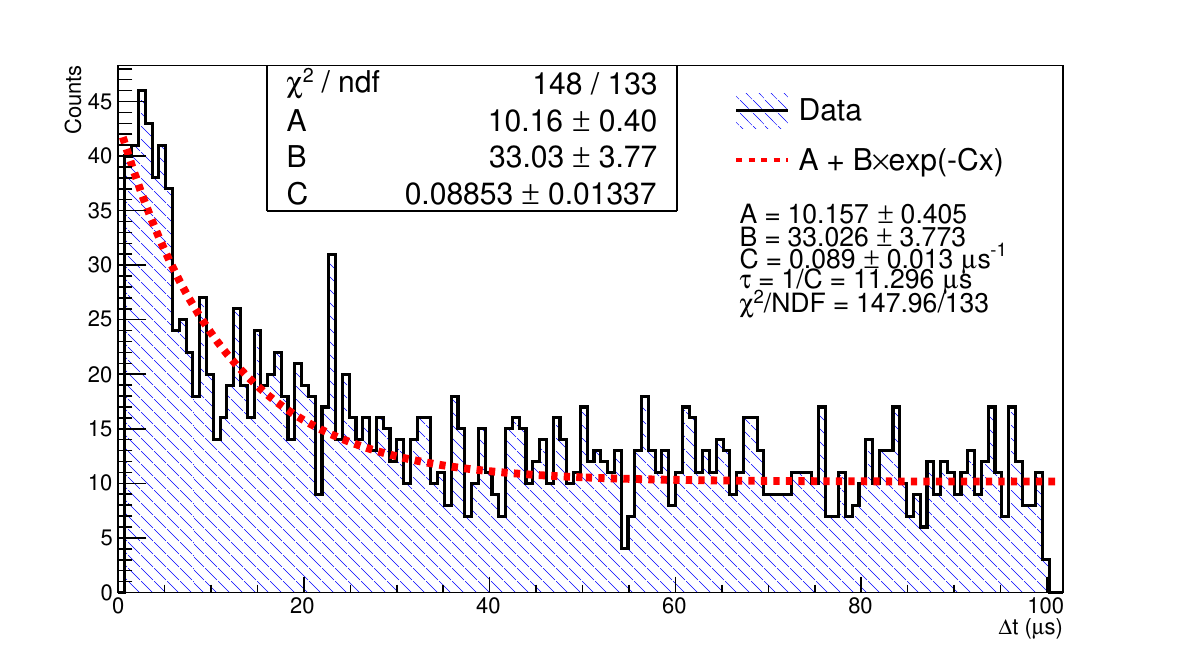}
    \caption{Left panel: Two-dimensional histogram of signal integral versus PSD values obtained with the SG101+EJ276 detector and acquired using the DT5751 digitizer, with events classified according to their PSD values; the right side lists the event categories and their respective counts. Right panel:The time differences of all correlated events within a 100-microsecond time window were calculated and fitted using the function form \( A + B \cdot e^{-Cx} \). The fitting results are: \( A = 10.157 \), \( B = 33.026 \), \( C = 0.08853 \, \mu\mathrm{s}^{-1} \), corresponding to a time constant \( \tau = 1/C = 11.296 \, \mu\mathrm{s} \).}
    \label{fig:corelative_event}
\end{figure}

Signals from the SG101+EJ276 detector system under irradiation by an Am--Be neutron source were acquired for ten minutes using a DT5751 digitizer, and the number of correlated events between fast neutrons in EJ276 and thermal neutrons in SG101 was statistically analyzed.
Figure~\ref{fig:corelative_event} left panel presents the PSD distribution from a ten-minute signal acquisition period. The DT5751 digitizer recorded a total of 635,687 events.Using PSD analysis, the selected events were classified into 316,708 gamma-ray events, 16,523 thermal neutron events, and 339,286 fast neutron events.The correlated events were selected by finding a fast neutron event in a 100\,\textmu s window prior to a thermal neutron event.

The accidental coincidence probability of a fast neutron and a thermal neutron is estimated as
\[
P = 1 - e^{-\lambda_f \tau}
= 1 - \exp\!\left(-\frac{339\,286}{600} \times 100 \times 10^{-6}\right) 
\approx 0.0550.
\]
The estimation is based on the Poisson distribution, where $\lambda_f$ denotes the average counting rate of fast neutrons and $\tau$ represents the coincidence time window.
The observed number of correlated events (2008) is significantly higher than the expected accidental coincidence events (approximately 908), indicating a genuine physical correlation was observed by the detector between thermal neutron events and fast neutron events occurring within the preceding 100\,\textmu s.In addition,events are selected in which a $\gamma$-ray signal is followed by a fast neutron signal within the 100\,\textmu s window preceding each thermal neutron event.A total of 108 triple-correlated events were obtained.The expected number of accidental triple coincidences with a $\gamma$-ray preceding a fast neutron within the time window $\tau$ is given by
\[
N_{\text{acc}} = N_t \cdot \lambda_\gamma \lambda_f \cdot \frac{\tau^2}{2}
= N_t \cdot \frac{N_\gamma}{T} \cdot \frac{N_f}{T} \cdot \frac{\tau^2}{2} \approx 25.
\]
where $N_t$, $N_\gamma$, and $N_f$ denote the total counts of thermal neutrons, $\gamma$-rays, and fast neutrons, respectively; $T$ is the total acquisition time; and $\tau$ is the coincidence time window 100\,\textmu s.The expected number of accidental coincidence events(25) is significantly lower than the observed count(108), indicating the presence of genuine triple-correlated events.

The right panel of Figure~\ref{fig:corelative_event} shows the distribution of time differences between fast and slow neutrons within a 100-microsecond time window.These data were fitted using a functional form \( A + B \cdot e^{-Cx} \), where \( x \) denotes the time difference. The best-fit parameters are \( A = 10.157 \), \( B = 33.026 \), and \( C = 0.08853 \, \mu\mathrm{s}^{-1} \). The exponential decay constant corresponds to a characteristic time constant \( \tau = 1/C = 11.296 \, \mu\mathrm{s} \), which reflects the typical timescale of the correlated event population under this SG101+EJ276 setup. This relatively short decay time can be an advantage in neutrino detector for inverse beta decay (IBD) event selection.

\section{Conclusions}
\label{sec:sum}

This work presents a systematic performance evaluation of the thermal neutron-sensitive transparent glass scintillator SG101 against the conventional LiF/ZnS:Ag-based scintillator EJ426. Under Am--Be neutron irradiation, SG101 exhibits higher detection efficiency: with identical experimental conditions, the event rate of SG101 is 6–8 times that of EJ426. In addition, the thermal neutron signal waveform of SG101 reveals significantly better signal stability, resulting in a better energy resolution.

The energy response linearity of composite detectors combining SG101 with organic scintillators (EJ200 or EJ276) was calibrated using standard $\gamma$ sources (\textsuperscript{22}Na, \textsuperscript{137}Cs, and \textsuperscript{60}Co), demonstrating excellent linearity in the 0.3--1.3\,MeV gamma-equivalent energy range.

Notably, the system shows outstanding particle discrimination capability: the SG101--EJ200 configuration achieves a FOM of 3.81(9.0$\sigma$) for thermal neutron/gamma separation, while the SG101--EJ276 setup clearly resolves three distinct particle populations—gamma rays, fast neutrons, and thermal neutrons—with FOM values of 3.46(8.1$\sigma$) for thermal neutron/$\gamma$ separation and 2.21(5.2$\sigma$) for thermal/fast neutron separation, respectively. Consequently, the thermal neutron signal from SG101 can be effectively distinguished from the fast signals produced by either organic scintillator at the $5\sigma$ confidence level.

Furthermore, coincidence analysis under Am--Be irradiation reveals a significant excess of genuine fast--thermal neutron events over the accidental background, confirming the system's practical suitability for applications requiring time-correlated event tagging in various nuclear detection field.The triple coincidence of SG101+EJ276 offers a potentially new technology for neutrino detection with highly background suppression such as fast neutrons and double neutrons, etc.

In summary, the SG101 transparent glass scintillator combines high optical transparency with high thermal neutron sensitivity. When coupled with organic scintillators capable of PSD, it forms a composite detection system capable of multi-particle discrimination and time-correlated event tagging. Its unique material properties and system integration advantages offer a practical pathway for developing compact neutron detectors with strong background suppression capabilities, demonstrating significant promise in advanced nuclear applications,neutrino detection, and radiation security.

\section*{Acknowledgments}
\label{sec:1:acknow}
This research was supported by the China Postdoctoral Science Foundation (Grant No. 2025M783462) , the Fundamental Research Funds for the Central Universities (Grant No. 24qnpy109) and the National Natural Science Foundation of China (Grant No. 12075087). We thank the School of Physics and Institut Franco-Chinois de l’Énergie Nucléaire at Sun Yat-sen University. We are also grateful to Prof. Yuehuan Wei, Prof. Bo Mei, and Prof. Tao Xiong for their support.

\printcredits

 \bibliographystyle{model1-num-names}

\bibliography{cas-refs}


\bio{}
Author biography without author photo.
Author biography. Author biography. Author biography.
\endbio

\end{document}